\documentclass[aps,prd,10pt,twocolumn,floatfix,superscriptaddress,preprintnumbers,tightenlines,showpacs,showkeys,nofootinbib,notoccite,notitlepage]{revtex4-1}
\usepackage[utf8]{inputenc}
\usepackage[colorlinks=true,citecolor=blue,linkcolor=blue]{hyperref}
\usepackage[normalem]{ulem}
\usepackage{amsmath,amssymb, mathrsfs}
\usepackage{epsfig}
\usepackage{graphicx}  
\usepackage{url}
\usepackage{color}
\usepackage{slashed}
\usepackage{multirow}
\usepackage{placeins}
\usepackage[dvipsnames]{xcolor}
\usepackage{epstopdf}
\usepackage{soul}
\usepackage{tikz}
\usepackage[capitalise, english]{cleveref}
\usepackage{siunitx}
\usepackage{xspace}
\usepackage{booktabs}
\usetikzlibrary{trees}
\usetikzlibrary{decorations.pathmorphing}
\usetikzlibrary{decorations.markings}

\newcommand{\appendixhead}%
{\begin{center}\textbf{\\Appendices\vspace{-0.5cm}}\end{center}}

\newcommand\myshade{80}
\colorlet{mylinkcolor}{ForestGreen}
\colorlet{mycitecolor}{Red}
\colorlet{myurlcolor}{violet}

\hypersetup{
	linkcolor  = mylinkcolor!\myshade!black,
	citecolor  = mycitecolor!\myshade!black,
	urlcolor   = myurlcolor!\myshade!black,
	colorlinks = true
}

\allowdisplaybreaks

\usepackage{tikz,xcolor,hyperref}

\definecolor{lime}{HTML}{A6CE39}
\DeclareRobustCommand{\orcidicon}{\hspace{-1mm}
	\begin{tikzpicture}
		\draw[lime, fill=lime] (0,0) 
		circle [radius=0.16] 
		node[white] {{\fontfamily{qag}\selectfont \tiny \,ID}};
		\draw[white, fill=white] (-0.0525,0.095) 
		circle [radius=0.007];
	\end{tikzpicture}
	\hspace{-3mm}
}

\foreach \x in {A, ..., Z}{\expandafter\xdef\csname orcid\x\endcsname{\noexpand\href{https://orcid.org/\csname orcidauthor\x\endcsname}
		{\noexpand\orcidicon}}
}

\begin{document}
	
	\title{Evolution of Tau-Neutrino Lepton Number in Protoneutron Stars due to Active-Sterile Neutrino Mixing}
	
	\author{Anupam Ray\orcidA{}}
	\email{anupam.ray@berkeley.edu}
	\affiliation{Department of Physics, University of California Berkeley, Berkeley, California 94720, USA}
	\affiliation{School of Physics and Astronomy, University of Minnesota, Minneapolis, MN 55455, USA}
	
	\author{Yong-Zhong Qian\orcidB{}}
	\email{qianx007@umn.edu}
	\affiliation{School of Physics and Astronomy, University of Minnesota, Minneapolis, MN 55455, USA}
	
	\date{\today}

	
	\begin{abstract}
		We present an approximate treatment of the mixing between $\nu_{\tau}$ ($\bar\nu_{\tau}$) and a sterile species 
		$\nu_s$ ($\bar\nu_s$) with a vacuum mass-squared difference of $\sim 10^2$--$10^3$~keV$^2$ in protoneutron stars created in core-collapse supernovae. Including production of sterile neutrinos through both resonant flavor conversion and collisions, we track the evolution of the $\nu_{\tau}$ lepton number due to both escape of sterile 
		neutrinos and diffusion. Our approach provides a reasonable treatment of the pertinent processes discussed in 
		previous studies and serves a pedagogical purpose to elucidate the relevant physics. We also discuss refinements 
		needed to study more accurately how flavor mixing with sterile neutrinos affects protoneutron star evolution.
	\end{abstract}
	
	\maketitle
	\preprint{N3AS-23-013}
	
	\section{Introduction}
	Sterile neutrinos associated with a vacuum mass eigenstate of $\mathcal{O}(10)$~keV in mass have been studied extensively as
	a viable dark matter candidate~\cite{Kusenko:2009up,Drewes:2016upu,Abazajian:2017tcc,Boyarsky:2018tvu,Dasgupta:2021ies}. Detection of an unidentified X-ray line near 3.55 keV by stacked observations of galaxy clusters and its possible explanation by the decay of sterile neutrinos~\cite{Bulbul:2014sua,Boyarsky:2014jta} have provided further stimulation for such studies. Among the exploration of the  effects of sterile neutrinos in cosmology and astrophysics, a number of studies have investigated their flavor mixing
	with active neutrinos in core-collapse supernovae~\cite{Shi:1993ee,Nunokawa:1997ct,Hidaka:2006sg,Hidaka:2007se,1993APh.....1..165R,2011PhRvD..83i3014R,Warren:2014qza,Warren:2016slz,2019PhRvD..99d3012A,2019JCAP...12..019S,Suliga:2020vpz,PhysRevD.106.015017}. In this paper, we revisit this topic to gain more insights into the underlying processes. Specifically, we focus on the mixing between $\nu_{\tau}$ 
	($\bar\nu_{\tau}$) and a sterile species $\nu_s$ ($\bar\nu_s$) with a vacuum mass-squared difference of 
	$\delta m^2\sim 10^2$--$10^3$~keV$^2$ in protoneutron stars created in core-collapse supernovae, and estimate the evolution of the $\nu_\tau$ lepton number as a result of such mixing. Our purpose is mainly pedagogical. While the key effects of $\nu_{\tau}$-$\nu_s$ and $\bar\nu_{\tau}$-$\bar\nu_s$ 
	mixing in protoneutron stars have been discussed by a number of previous studies (e.g., 
	\cite{1993APh.....1..165R,2011PhRvD..83i3014R,2019PhRvD..99d3012A,2019JCAP...12..019S,PhysRevD.106.015017}), 
	we wish to elucidate the underlying physics by examining various aspects of the problem, including assumptions 
	and approximations made to facilitate a reasonable treatment.
	
	The following previous studies are especially pertinent to our discussion. In Ref.~\cite{1993APh.....1..165R}, the formalism for treating production of sterile neutrinos through collisions of active neutrinos in the supernova 
	core was developed. Ref.~\cite{2011PhRvD..83i3014R} pointed out that escape of $\nu_s$ ($\bar\nu_s$) mixed with $\nu_\tau$ 
	($\bar\nu_\tau$) leads to evolution of the $\nu_\tau$ lepton number, which tends to turn off the effects of such 
	flavor mixing. Ref.~\cite{2019PhRvD..99d3012A} highlighted that
	$\bar\nu_{\tau}$-$\bar\nu_s$ conversion is greatly enhanced by the Mikheyev–Smirnov–Wolfenstein (MSW) effect 
	\cite{MikheyeSmirnov1985,1978PhRvD..17.2369W} when the neutrino mean free path exceeds the width of the resonance region.
	A careful numerical study of $\nu_{\tau}$-$\nu_s$ and $\bar\nu_{\tau}$-$\bar\nu_s$ mixing was carried out in
	Ref.~\cite{2019JCAP...12..019S} with emphasis on the feedback of such mixing. Finally, the effects of diffusion of the
	$\nu_\tau$ lepton number created by such mixing were discussed in Ref.~\cite{PhysRevD.106.015017}. However, a concise and analytical treatment of all the relevant processes appears lacking, and we aim to present such an approach here.
	
	The rest of the paper is organized as follows. In Section~\ref{sec:setup} we set up the problem of $\nu_{\tau}$-$\nu_s$ and $\bar\nu_{\tau}$-$\bar\nu_s$ mixing with
	$\delta m^2\sim 10^2$--$10^3$~keV$^2$ in protoneutron stars. We discuss the production of sterile neutrinos through
	both the MSW effect and collisions, as well as the diffusion of the $\nu_\tau$ lepton number created by such mixing.  
	In Section~\ref{sec:example} we present example calculations to illustrate the overall evolution of the $\nu_{\tau}$ lepton number in protoneutron stars due to
	both escape of sterile neutrinos and diffusion.
	In Section~\ref{sec:feedback} we elaborate on the feedback of $\nu_{\tau}$-$\nu_s$ and $\bar\nu_{\tau}$-$\bar\nu_s$ mixing by focusing on the evolution of
	the $\nu_{\tau}$ lepton number in a specific radial zone. In Section~\ref{sec:discuss}
	we give conclusions and discuss refinements needed to study more accurately how flavor mixing with sterile neutrinos 
	affects protoneutron star evolution.
	
	\section{Active-sterile neutrino mixing in protoneutron stars}
	\label{sec:setup}
	We consider the mixing of $\nu_{\tau}$ ($\bar\nu_{\tau}$) and $\nu_s$ ($\bar\nu_s$) with $\delta m^2\sim 10^2$--$10^3$~keV$^2$
	and vacuum mixing angle $\theta\ll 1$. In protoneutron stars, forward scattering on electrons, protons, neutrons, and neutrinos,
	whose number densities are assumed to depend on radius only, results in a potential 
	\begin{align}
		V_{\nu}=\sqrt{2}G_Fn_b\left[-\frac{1-Y_e}{2}+Y_{\nu_e}+Y_{\nu_\mu}+2Y_{\nu_\tau}\right]
	\end{align}
	for $\nu_{\tau}$-$\nu_s$ mixing, where $G_F$ is the Fermi constant, $n_b$ is the baryon number density, $Y_e$ is the electron 
	fraction (net number of electrons per baryon), $Y_{\nu_\alpha}=(n_{\nu_\alpha}-n_{\bar\nu_\alpha})/n_b$ with $\alpha=e$, $\mu$,
	and $\tau$ is the $\nu_\alpha$ lepton number fraction, and $n_{\nu_\alpha}$ ($n_{\bar\nu_\alpha}$) is the number density of $\nu_\alpha$ 
	($\bar\nu_\alpha$). Note that $n_b$, $Y_e$, $Y_{\nu_e}$, $Y_{\nu_\mu}$, $Y_{\nu_\tau}$, and hence $V_\nu$ are all functions of radius, but for convenience, here and below we usually suppress such radial dependence.
	The corresponding potential for $\bar\nu_{\tau}$-$\bar\nu_s$ mixing is $V_{\bar\nu}=-V_{\nu}$.
	The above potentials modify the $\nu_{\tau}$-$\nu_s$ and $\bar\nu_{\tau}$-$\bar\nu_s$ mixing in protoneutron stars, which is
	characterized by the effective mixing angles $\theta_\nu$ and $\theta_{\bar\nu}$, respectively. For a neutrino with energy $E$,
	\begin{align}
		\sin^22\theta_\nu&=\frac{\Delta^2\sin^22\theta}{(\Delta\cos2\theta-V_{\nu})^2+\Delta^2\sin^22\theta},\label{eq:thetanu}\\
		\sin^22\theta_{\bar\nu}&=\frac{\Delta^2\sin^22\theta}{(\Delta\cos2\theta-V_{\bar\nu})^2+\Delta^2\sin^22\theta},\label{eq:thetanubar}
	\end{align}
	where $\Delta=\delta m^2/(2E)$.
	
	For conditions in protoneutron stars, $V_\nu<0$ and $V_{\bar\nu}>0$, so $\nu_{\tau}$-$\nu_s$ mixing is suppressed but there is
	a resonance for $\bar\nu_{\tau}$-$\bar\nu_s$ mixing when $\Delta\cos2\theta=V_{\bar\nu}$. For $\theta\ll1$, the resonance energy 
	$E_R$ corresponding to a specific value of $V_{\bar\nu}$ is
	\begin{align}
		E_R=\frac{\delta m^2}{2V_{\bar\nu}}=\frac{13.1\ {\rm MeV}(\delta m^2/10^2\ {\rm keV}^2)}
		{(1-Y_e-2Y_{\nu_e}-2Y_{\nu_\mu}-4Y_{\nu_\tau})\rho_{14}},
	\end{align}
	where $\rho_{14}$ is the mass density in units of $10^{14}$~g~cm$^{-3}$.
	By definition, the resonance region for a $\bar\nu_{\tau}$ with energy $E_R$ corresponds to $\sin^22\theta_{\bar\nu}\geq 1/2$.
	For a radially-propagating $\bar\nu_{\tau}$, this region has a radial width
	\begin{align}
		\delta r=2H_R\tan2\theta,
	\end{align}
	where $H_R=|\partial\ln V_{\bar\nu}/\partial r|_{E_R}^{-1}$ and the derivative is taken at the resonance radius for $E_R$. For $\theta\ll1$,
	we can make the approximation that $\theta_{\bar\nu}$ changes from $\pi/2$ to 0 (or vice versa) once the resonance region is 
	traversed and use the Landau-Zener formula (e.g., \cite{1986PhRvL..57.1271H,1986PhRvL..57.1275P}) to estimate the survival
	probability for a radially-propagating $\bar\nu_\tau$ as
	\begin{align}\label{eq:LZ}
		P_{\rm LZ}=\exp\left(-\frac{\pi^2\delta r}{2L_{\rm res}}\right)=e^{-\gamma_R},
	\end{align}
	where 
	\begin{align}
		L_{\rm res}=\frac{4\pi E_R}{\delta m^2\sin2\theta}
	\end{align}
	is the oscillation length at resonance and
	\begin{align}\label{eq:gamma}
		\gamma_R=39.8\left(\frac{\sin^22\theta}{10^{-10}}\right)\left(\frac{\delta m^2}{10^2\ {\rm keV}^2}\right)
		\left(\frac{\rm MeV}{E_R}\right)
		\left(\frac{H_R}{\rm km}\right)
	\end{align}
	is the adiabaticity parameter. For $\gamma_R\gg1$, flavor evolution is highly adiabatic with $P_{\rm LZ}\sim 0$
	and $\bar\nu_\tau$ is efficiently converted into $\bar\nu_s$.
	
	Description of $\nu_s$ and $\bar\nu_s$ production requires knowledge about $\nu_\tau$ and $\bar\nu_\tau$ in protoneutron stars.
	We focus on the region where all active neutrinos are diffusing and assume that they are always in local thermodynamic equilibrium 
	with the matter background (i.e., neutrons, protons, electrons, positrons, and photons). Consequently, the $\nu_\alpha$ and
	$\bar\nu_\alpha$ energy distributions (number densities per unit energy interval per unit solid angle) are
	\begin{align}
		\frac{d^2n_{\nu_\alpha}}{dEd\Omega}&=\frac{1}{(2\pi)^3}\frac{E^2}{e^{(E-\mu_{\nu_\alpha})/T}+1},\label{eq:dnudedomega}\\
		\frac{d^2n_{\bar\nu_\alpha}}{dEd\Omega}&=\frac{1}{(2\pi)^3}\frac{E^2}{e^{(E+\mu_{\nu_\alpha})/T}+1},\label{eq:dnubardedomega}
	\end{align}
	where $d\Omega$ is the differential solid angle in the direction of the momentum, $\mu_{\nu_\alpha}$ is the $\nu_\alpha$ chemical
	potential, and $T$ is the matter temperature. The $\nu_\alpha$ lepton number fraction is
	\begin{align}
		Y_{\nu_\alpha}=\frac{T^3\eta_{\nu_\alpha}}{6n_b}\left(1+\frac{\eta_{\nu_\alpha}^2}{\pi^2}\right),
	\end{align}
	where $\eta_{\nu_\alpha}=\mu_{\nu_\alpha}/T$. Protoneutron stars have $Y_{\nu_\mu}\ll Y_{\nu_e}\ll 1-Y_e$ and we assume 
	$Y_{\nu_\mu}=0$ hereafter. Because we only consider $\nu_{\tau}$-$\nu_s$ and $\bar\nu_{\tau}$-$\bar\nu_s$ mixing, neither 
	$Y_{\nu_e}$ nor $Y_{\nu_\mu}$ is affected by such mixing, but the feedback on $Y_{\nu_\tau}$ is critical to our discussion.
	
\begin{figure*}[!t]
		\centering
		\includegraphics[width=0.245\textwidth]{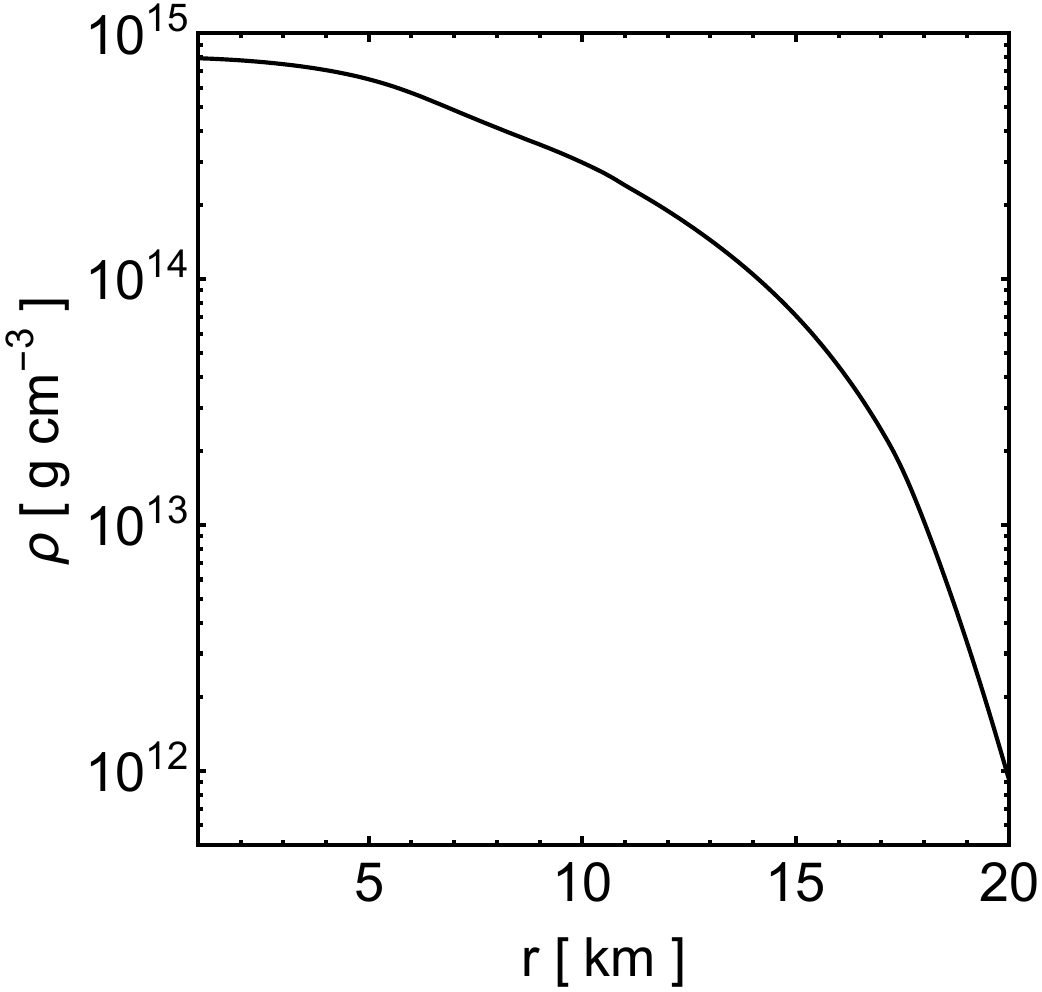}	
		\hspace{0.1 cm}
		\includegraphics[width=0.23\textwidth]{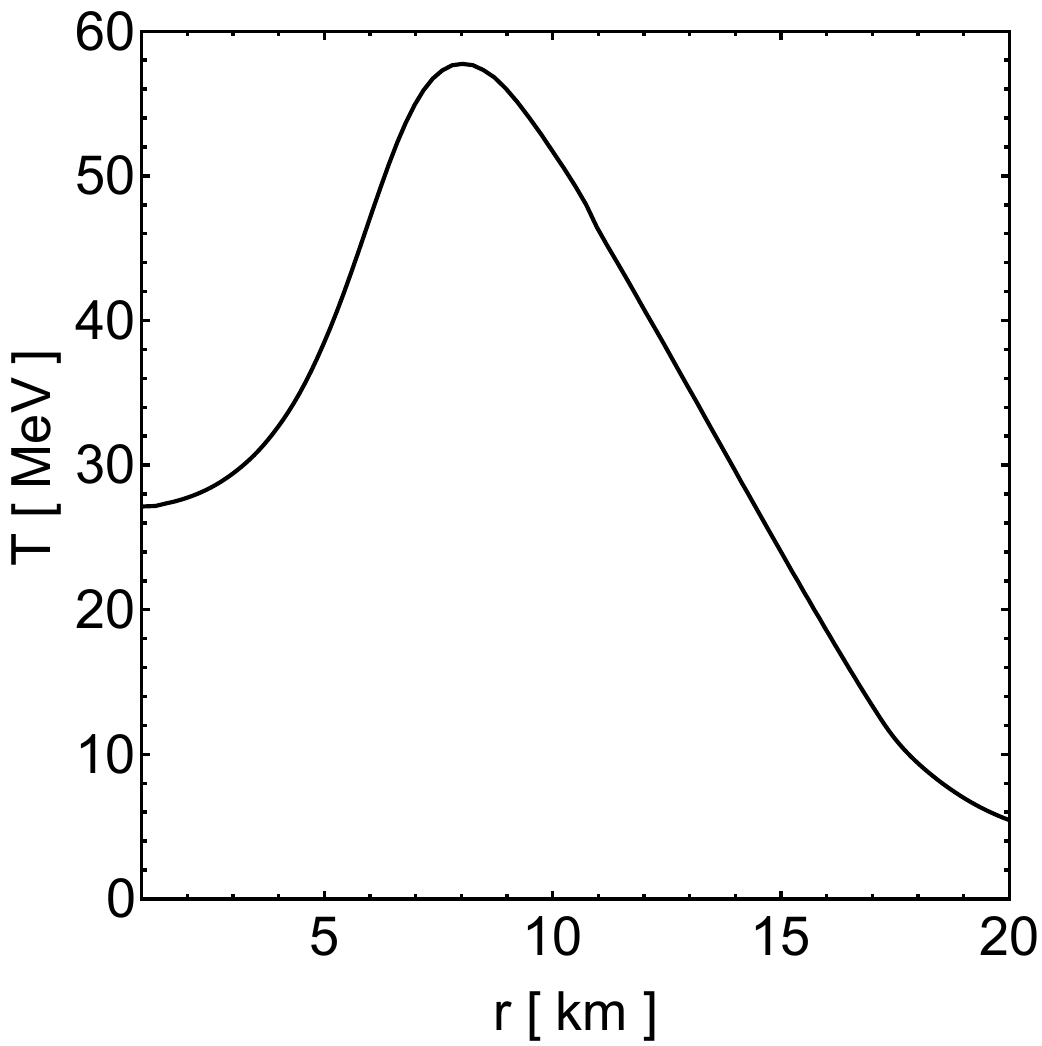}
		\hspace{0.1 cm}
		\includegraphics[width=0.24\textwidth]{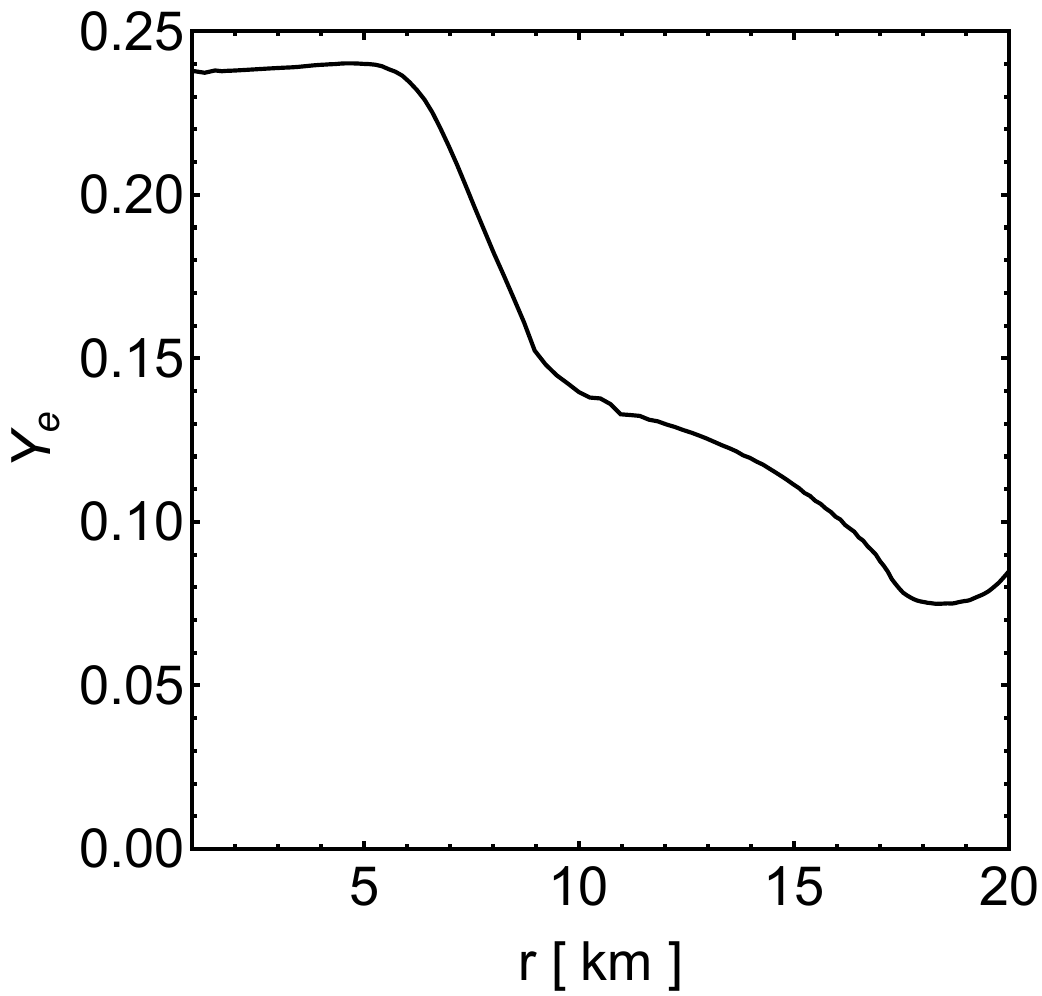}	
		\hspace*{0.0 cm}
		\includegraphics[width=0.24\textwidth]{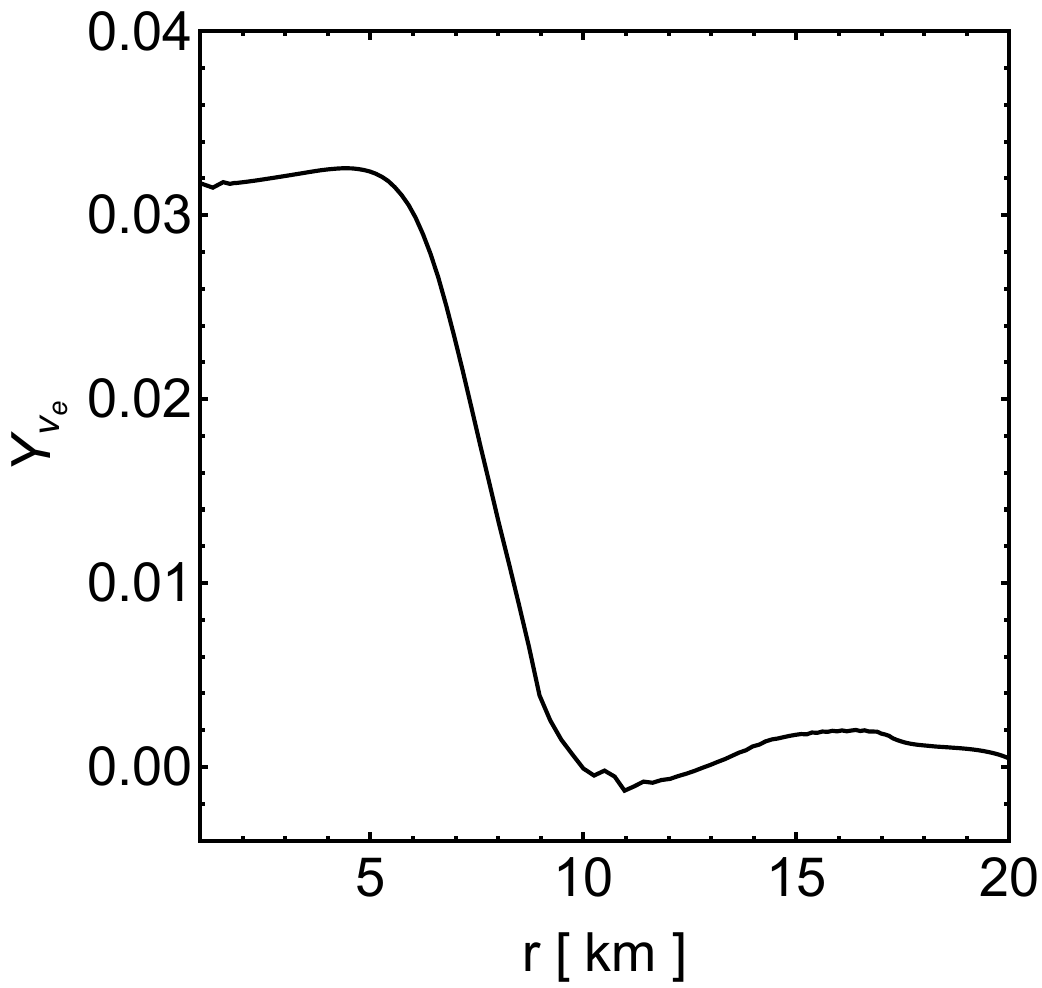}
		\caption{Protoneutron star conditions from a $20\,M_\odot$ supernova model (with the SFHo nuclear equation of state) at 1 s post core bounce~\cite{Bollig2016}. 
			The radial profiles of $\rho$, $T$, $Y_e$, and $Y_{\nu_e}$ are shown.}
		\label{fig:pns}
	\end{figure*}

	For numerical examples, we use the protoneutron star conditions from a $20\,M_\odot$ supernova model (with the SFHo nuclear equation of state)
	at 1 s post core bounce~\cite{Bollig2016,Mirizzi:2015eza,Bollig:2017lki,Bollig:2020xdr}. The corresponding radial profiles of $\rho$, $T$, $Y_e$, and $Y_{\nu_e}$ 
	are shown in Fig.~\ref{fig:pns}.
	
	\subsection{Sterile neutrino production through MSW effect}
	\label{sec:MSW}
Inside a protoneutron star, while active neutrinos are diffusing, sterile neutrinos produced through $\nu_\tau$-$\nu_s$ and $\bar\nu_\tau$-$\bar\nu_s$ mixing readily pass through matter without interaction, thereby affecting the $\nu_\tau$ lepton number distribution.
	For a homogeneous medium with uniform density and temperature, sterile neutrinos are produced only when active
	neutrinos collide with medium particles and the problem was treated in Ref.~\cite{2011PhRvD..83i3014R}. For realistic protoneutron stars, both
	density and temperature vary with radius (see Fig.~\ref{fig:pns}). Ref.~\cite{2019PhRvD..99d3012A} highlighted that $\bar\nu_\tau$ can be resonantly 
	converted into $\bar\nu_s$ through the MSW effect
	when the $\bar\nu_\tau$ mean free path is longer than the size of the resonance region $\delta r$. The mean free path is
	determined predominantly by neutral-current scattering on neutrons and protons for both $\nu_\tau$ and $\bar\nu_\tau$. 
	It can be estimated as
	\begin{align}
		\lambda(E)=\frac{1}{n_b\sigma(E)}\approx\frac{9.85\ {\rm km}}{\rho_{14}}\left(\frac{\rm MeV}{E}\right)^2,
	\end{align}
	where $\sigma(E)\approx G_F^2E^2/\pi$ is the cross section (e.g., \cite{2011PhRvD..83i3014R}). 
	
	Denoting $\lambda_R=\lambda(E_R)$, we consider a volume element of linear size $l\sim\lambda_R>\delta r$ 
	that surrounds a point on the resonance sphere for energy $E_R$. Inside this volume element, $\bar\nu_\tau$ with energy close 
	to $E_R$ and propagating radially outward can cross their 
	resonance regions without encountering collisions and be converted into $\bar\nu_s$ with a probability of $1-P_{\rm LZ}$. These $\bar\nu_s$ then escape from the protoneutron star.
	Specifically, the energy interval of such $\bar\nu_\tau$ is
	\begin{align}
		\delta E=\left|\frac{\partial E_R}{\partial r}\right|l=\frac{l}{H_R}E_R.
	\end{align}
	However, the angular distribution of $\bar\nu_\tau$ is isotropic [see Eq.~(\ref{eq:dnubardedomega})].
	A $\bar\nu_\tau$ propagating with a polar angle $\vartheta\sim\pi/2$ relative to the radially outward direction experiences 
	less change of $V_{\bar\nu}$ inside the volume element than a radially-propagating $\bar\nu_\tau$ and therefore, 
	may fail to traverse its full resonance region before encountering collisions. In this case, the MSW effect would be 
	reduced significantly because flavor evolution now depends on the specific values of $V_{\bar\nu}$ between successive collisions
	(i.e., we can no longer assume that $\theta_{\bar\nu}$ changes from $\pi/2$ to 0 or vice versa in crossing the resonance).
	In addition, a $\bar\nu_\tau$ propagating radially inward has a probability of $1-P_{\rm LZ}$ to be converted
		into a $\bar\nu_s$ after going through the first resonance, and this $\bar\nu_s$ readily passes through the interior of the
		resonance sphere to encounter a second resonance on the opposite side of the center. The probability for this $\bar\nu_s$ to remain as a 
		$\bar\nu_s$ and escape from the protoneutron star is $P_{\rm LZ}$.
	
	While a full treatment should account for all the directional dependence of $\bar\nu_s$ production through the MSW effect,
	we provide an approximate treatment by considering the effective flux $\Phi_{\bar\nu_s}^{\rm MSW}$ of $\bar\nu_s$ 
	(number per unit area per unit time) escaping radially 
	(both outward and inward) from the volume element. Because the contribution to this flux is weighed by $\cos\vartheta$,
	we assume that $\bar\nu_\tau$ with $0\leq\vartheta<\pi/2$ ($\pi/2<\vartheta\leq\pi$) experience the same MSW effect as
	a $\bar\nu_\tau$ propagating radially outward (inward). Taking into account that a $\bar\nu_\tau$ propagating radially
	inward goes through two resonance regions, we obtain
	\begin{align}
		\Phi_{\bar\nu_s}^{\rm MSW}&=\Theta(\lambda_R-\delta r)(1-P_{\rm LZ})
		\left.\frac{d^2n_{\bar\nu_\tau}}{dEd\Omega}\right|_{E_R}\delta E\nonumber\\
		&\times\left[\int_{\rm out}d\Omega\,\cos\vartheta+P_{\rm LZ}\int_{\rm in}d\Omega\,|\cos\vartheta|\right]\\
		&=\Theta(\lambda_R-\delta r)\pi(1-P_{\rm LZ}^2)\left.\frac{d^2n_{\bar\nu_\tau}}{dEd\Omega}\right|_{E_R}\delta E,
	\end{align}
	where $\Theta(x)$ is the Heaviside step function. Because $\bar\nu_s$ readily escape from the protoneutron star, the $\nu_\tau$ lepton number in the volume element is increased by the conversion of $\bar\nu_\tau$ into $\bar\nu_s$. So we estimate that due to the MSW effect, the rate of change of $Y_{\nu_\tau}$ 
	in the volume element is
	\begin{align}
		\dot Y_{\nu_\tau}^{\rm MSW}&=\frac{\Phi_{\bar\nu_s}^{\rm MSW}}{n_bl},\\
		&=\Theta(\lambda_R-\delta r)
		\frac{\pi E_R(1-P_{\rm LZ}^2)}{n_bH_R}\left.\frac{d^2n_{\bar\nu_\tau}}{dEd\Omega}\right|_{E_R},\label{eq:rymsw}\\
		&=\frac{222\ {\rm s}^{-1}}{\rho_{14}}\left(\frac{E_R}{\rm 30\ MeV}\right)^3\left(\frac{\rm km}{H_R}\right)\nonumber\\
		&\times\frac{\Theta(\lambda_R-\delta r)(1-P_{\rm LZ}^2)}{e^{(E_R/T)+\eta_{\nu_\tau}}+1}.\label{eq:rymsw2}
	\end{align}
	
	The above results are derived here for the first time, although they are qualitatively similar to those assumed in 
	Refs.~\cite{2019PhRvD..99d3012A,2019JCAP...12..019S,PhysRevD.106.015017}. Quantitatively, the rate of change of 
	$Y_{\nu_\tau}$ due to the MSW effect used in the previous studies is approximately
	\begin{align}\label{eq:oldrymsw}
		\dot Y_{\nu_\tau}^{\rm MSW'}=\Theta(\lambda_R-\delta r)\frac{4\pi E_R(1-P_{\rm LZ})}{n_bH_R}
		\left.\frac{d^2n_{\bar\nu_\tau}}{dEd\Omega}\right|_{E_R},
	\end{align}
	which assumes that $\bar\nu_\tau$ experience the same MSW effect regardless of their direction of propagation and
	therefore, is clearly an overestimate. While $\dot Y_{\nu_\tau}^{\rm MSW}$ might be a more realistic estimate, we will 
	present results using both $\dot Y_{\nu_\tau}^{\rm MSW}$ and $\dot Y_{\nu_\tau}^{\rm MSW'}$ for comparison.
	
	\subsection{Sterile neutrino production through collisions}
Apart from the resonant MSW production, $\nu_s$ ($\bar\nu_s$) can also be produced via collisions of $\nu_\tau$ ($\bar\nu_\tau$) with the protoneutron star constituents. In this scenario, $\nu_\tau$-$\nu_s$ ($\bar\nu_\tau$-$\bar\nu_s$) mixing with the effective mixing angle $\theta_\nu$ ($\bar\theta_\nu$) allows a $\nu_\tau$ ($\bar\nu_\tau$) to evolve as a linear combination of two effective mass eigenstates between collisions. Upon collision, which is predominantly with neutrons and protons, the wave function collapses, thereby producing a $\nu_s$ ($\bar\nu_s$) with a probability proportional to $\sin^22\theta_\nu$ ($\sin^22\theta_{\bar\nu}$). Note that under the assumption of local thermodynamic equilibrium between active neutrinos and matter, emission and absorption processes mainly serve to produce the Fermi-Dirac distributions for $\nu_\tau$ and $\bar\nu_\tau$ 
		[see Eqs.~(\ref{eq:dnudedomega}) and (\ref{eq:dnubardedomega})]. Because collisions of $\nu_\tau$ and $\bar\nu_\tau$ with neutrons and protons 
		have much higher rates than their absorption processes, we average over their flavor evolution between collisions to obtain the effective rates
		for producing $\nu_s$ and $\bar\nu_s$.
	
\begin{figure}[!t]
		\centering
		\hspace*{-0.8 cm}
		\includegraphics[width=0.45\textwidth]{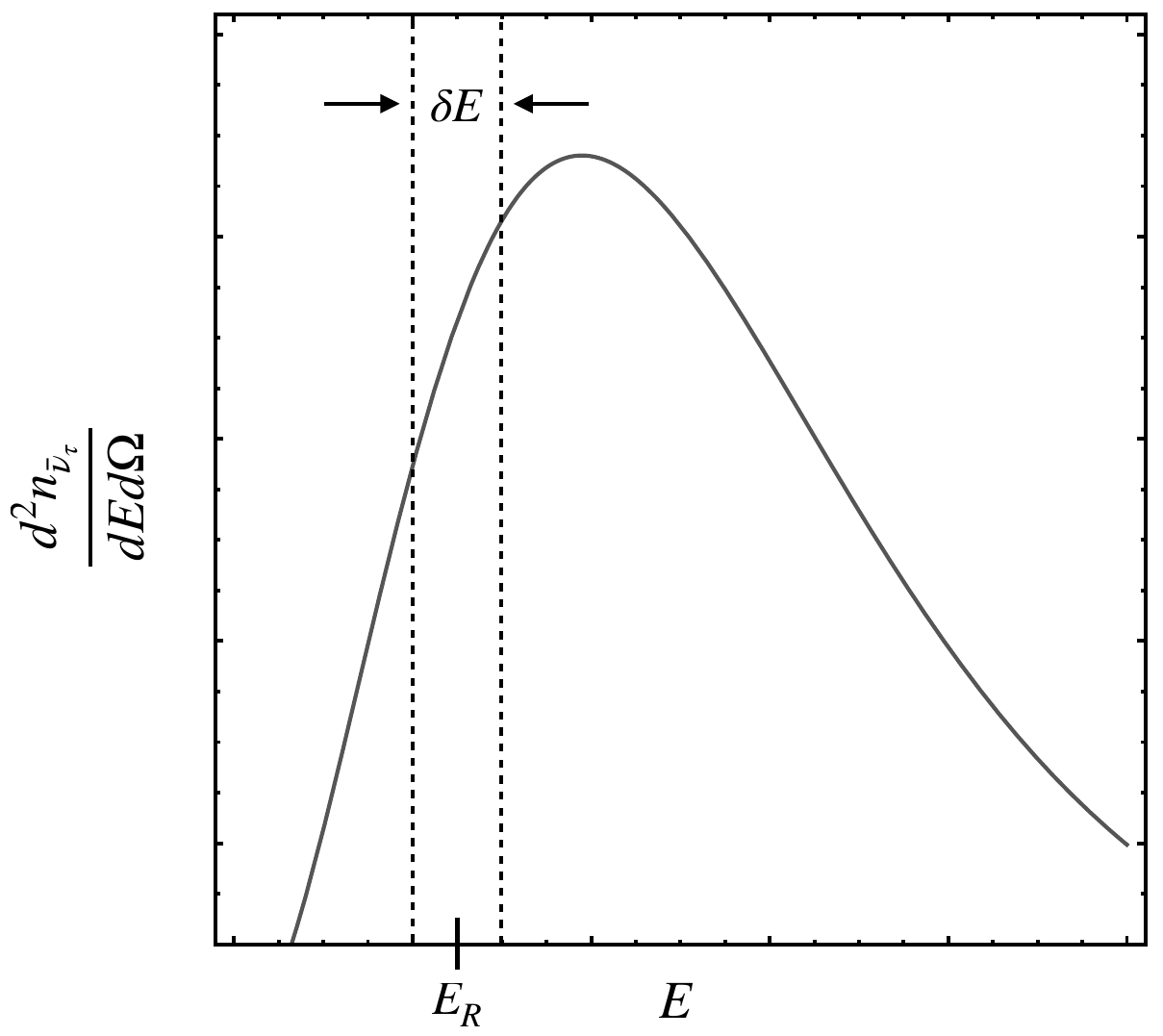}	
		\caption{Sketch of the $\bar\nu_\tau$ energy distribution at a specific radius. This distribution is determined by the $T$ and $\mu_{\nu_\tau}$ at this radius. For a given set of mixing parameters, the potential $V_{\bar\nu}$ at this radius defines a resonance energy $E_R$. Inside a volume element of linear size $\lambda_R$ centered at this radius, $\bar\nu_\tau$ with energy $[E_R-\delta E/2,E_R+\delta E/2]$ go through the MSW resonances. Here $\lambda_R$ is the mean free path for $\bar\nu_\tau$ with energy $E_R$ and $\delta E=(\lambda_R/H_R)E_R$ with $H_R=|\partial\ln V_{\bar\nu}/\partial r|_{E_R}^{-1}$.}
		\label{fig:spec}
	\end{figure}

	In the same volume element considered in Section~\ref{sec:MSW},
	the resulting rate of change of $Y_{\nu_\tau}$ due to flavor evolution and collisions is (e.g., \cite{1993APh.....1..165R})
	\begin{align}
		\dot Y_{\nu_\tau}^{\rm coll}&=\frac{\pi}{n_b}\left[\int_{E_{\bar\nu_\tau}}dE\,\frac{\sin^22\theta_{\bar\nu}}{\lambda(E)}
		\frac{d^2n_{\bar\nu_\tau}}{dEd\Omega}\right.\nonumber\\
		&\left.-\int_0^\infty dE\,\frac{\sin^22\theta_{\nu}}{\lambda(E)}\frac{d^2n_{\nu_\tau}}{dEd\Omega}\right],\\
		&=G_F^2\left(\int_{E_{\bar\nu_\tau}}dE\,E^2\sin^22\theta_{\bar\nu}
		\frac{d^2n_{\bar\nu_\tau}}{dEd\Omega}\right.\nonumber\\
		&\left.-\int_0^\infty dE\,E^2\sin^22\theta_{\nu}\frac{d^2n_{\nu_\tau}}{dEd\Omega}\right),
	\end{align}
	where the integration over $E_{\bar\nu_\tau}$ excludes the range $[E_R-\delta E/2,E_R+\delta E/2]$
	with $\delta E=(\lambda_R/H_R)E_R$ for resonant $\bar\nu_\tau$ (see Fig.~\ref{fig:spec})
	when $\lambda_R$ exceeds $\delta r$ and goes from 0 to $\infty$ otherwise. The above rate is similar to that used in Ref.~\cite{2011PhRvD..83i3014R},
except for the treatment of resonant $\bar\nu_\tau$ with $\lambda_R>\delta r$. This difference comes about because Ref.~\cite{2011PhRvD..83i3014R} approximated the protoneutron star as a uniform medium for which $\nu_s$ and $\bar\nu_s$ are produced
		with effective rates of $\sin^22\theta_\nu/[4\lambda(E)]$ and $\sin^22\theta_{\bar\nu}/[4\lambda(E)]$ by collisions of $\nu_\tau$ and $\bar\nu_\tau$, respectively. In contrast, for the realistic and radially-varying protoneutron star conditions adopted
		here, resonant $\bar\nu_\tau$ with $\lambda_R>\delta r$ are converted into $\bar\nu_s$ through the MSW effect in a nonuniform medium as discussed in Sec.~\ref{sec:MSW}, while the above effective rates still apply to $\nu_s$ production by collisions of $\nu_\tau$ and to $\bar\nu_s$ production by collisions of nonresonant $\bar\nu_\tau$ and by collisions of resonant $\bar\nu_\tau$ when $\lambda_R$ falls below $\delta r$.
	
Because $\sin^22\theta_{\bar\nu}$ sharply peaks at $E=E_R$ for $\theta\ll1$ [see Eq.~(\ref{eq:thetanubar}) and note that 
		$V_{\bar\nu}>0$] and $\int_0^\infty dE\sin^22\theta_{\bar\nu}\approx\pi E_R\tan2\theta$, we can make the approximation
		$\sin^22\theta_{\bar\nu}\approx\pi E_R\delta(E-E_R)\tan2\theta$. Integrating over the full range of $E_{\bar\nu_\tau}$ 
		(i.e., ignoring the exclusion of resonant $\bar\nu_\tau$ with $\lambda_R>\delta r$),
	we estimate an upper limit for the contribution to $\dot Y_{\nu_\tau}^{\rm coll}$ from $\bar\nu_\tau$-$\bar\nu_s$ mixing as follows: 
	\begin{align}
		\dot Y_{\nu_\tau}^{{\rm coll,\bar\nu}}\lesssim\frac{0.6\ {\rm s}^{-1}}{e^{(E_R/T)+\eta_{\nu_\tau}}+1}
		\left(\frac{E_R}{30\ {\rm MeV}}\right)^5\left(\frac{\sin2\theta}{10^{-5}}\right),
		\label{eq:rycnb}
	\end{align}
	where we have used $\tan2\theta\approx\sin2\theta$ to a very good approximation.
	Because $\sin^22\theta_\nu<\sin^22\theta$ [see Eq.~(\ref{eq:thetanu}) and note that $V_\nu<0$], we replace $\sin^22\theta_\nu$ with $\sin^22\theta$ in the corresponding integral
		to obtain an upper limit for the contribution to $\dot Y_{\nu_\tau}^{\rm coll}$ from $\nu_\tau$-$\nu_s$ mixing as follows:
	\begin{align}
		|\dot Y_{\nu_\tau}^{{\rm coll},\nu}|&<2\times10^{-6}\ {\rm s}^{-1}\left(\frac{T}{30\ {\rm MeV}}\right)^5\left(\frac{\sin^22\theta}{10^{-10}}\right)F_4(\eta_{\nu_\tau}),
		\label{eq:rycn}
	\end{align}
	where $F_4(\eta_{\nu_\tau})$ is the Fermi-Dirac integral of rank 4 and argument $\eta_{\nu_\tau}$. In general,
	\begin{align}
		F_k(\eta)=\int_0^\infty dx\ \frac{x^k}{e^{x-\eta}+1}.
	\end{align}
	Comparing Eqs.~(\ref{eq:rycnb}) and (\ref{eq:rycn}) with (\ref{eq:rymsw2}), we expect that the rate of change of $Y_{\nu_\tau}$ is dominated by $\dot Y_{\nu_\tau}^{\rm MSW}$ when $\lambda_R$ exceeds $\delta r$ and that this rate decreases dramatically
	to $\dot Y_{\nu_\tau}^{{\rm coll}}$ when $\lambda_R$ falls below $\delta r$. 
	
	\subsection{Diffusion of tau-neutrino lepton number}
	\label{sec:diff}
	As we focus on the region where all active neutrinos are diffusing, the outer radius $R_\nu$ of this region is an important
	quantity. For a $\nu_\tau$ or $\bar\nu_\tau$ of energy $E$, the transition from diffusion to free-streaming occurs at radius
	$r$ for which $\int_r^{\infty}dr/\lambda(E)\sim 1$. Considering the energy distributions of $\nu_\tau$ and $\bar\nu_\tau$, we estimate $R_\nu$ to be given by
	\begin{align}
		\int_{R_\nu}^{\infty}\frac{dr}{\langle\lambda(E)\rangle_{\bar\nu_\tau}}=\frac{G_F^2}{\pi}\int_{R_\nu}^{\infty}dr\,n_bT^2\frac{F_2(-\eta_{\nu_\tau})}{F_0(-\eta_{\nu_\tau})}=1,
	\end{align}
	where $\langle\lambda(E)\rangle_{\bar\nu_\tau}$ is the mean free path averaged over the $\bar\nu_\tau$ energy distribution. The use of $\langle\lambda(E)\rangle_{\bar\nu_\tau}$ gives a more stringent estimate of $R_\nu$, which typically corresponds to a density $\rho\sim 10^{13}$~g~cm$^{-3}$.
	
	The net $\nu_\tau$ lepton number flux crossing a surface at radius $r$ in the radially outward direction is
	\begin{align}
		\Phi_{\nu_\tau}(r)&=\int_0^\infty dE\,\frac{\lambda(E)}{3}\frac{\partial}{\partial r}\left(\frac{dn_{\bar\nu_{\tau}}}{dE}-\frac{dn_{\nu_\tau}}{dE}\right),\\
		&=-\frac{1}{6\pi G_F^2n_b}\frac{\partial\mu_{\nu_\tau}}{\partial r}.
	\end{align}
We have checked that $|\Phi_{\nu_\tau}(r)|$ is always below the product of $n_bY_{\nu_\tau}$ and the speed of light
		in the diffusion region of $r<R_\nu$. Therefore, the important causality limit on the diffusive flux (e.g., \cite{1981ApJ...248..321L}) is satisfied.
	
	Due to the above diffusive flux, the rate of change of $Y_{\nu_\tau}$ in the volume element considered in Section~\ref{sec:MSW} is
	\begin{align}
		\dot Y_{\nu_\tau}^{\rm diff}=\frac{1}{6\pi G_F^2n_br^2}\frac{\partial}{\partial r}\left(\frac{r^2}{n_b}\frac{\partial\mu_{\nu_\tau}}{\partial r}\right).
	\end{align}
	This result is the same as that used in Ref.~\cite{PhysRevD.106.015017} when the typo in its Eq.~(10) is corrected.
	As a crude estimate, the above rate is of the order $\sim 0.1\ {\rm s}^{-1}\rho_{14}^{-2}(\partial^2\mu_{\nu_\tau,{\rm MeV}}/\partial r_{\rm km}^2)$,
	where $\mu_{\nu_\tau,{\rm MeV}}$ is $\mu_{\nu_\tau}$ in units of MeV and $r_{\rm km}$ is $r$ in units of km. We expect that this rate
	is also small compared to $\dot Y_{\nu_\tau}^{\rm MSW}$ when $\lambda_R$ exceeds $\delta r$.
	
	\begin{figure*}[!t]
		\centering
		\includegraphics[width=0.3\textwidth]{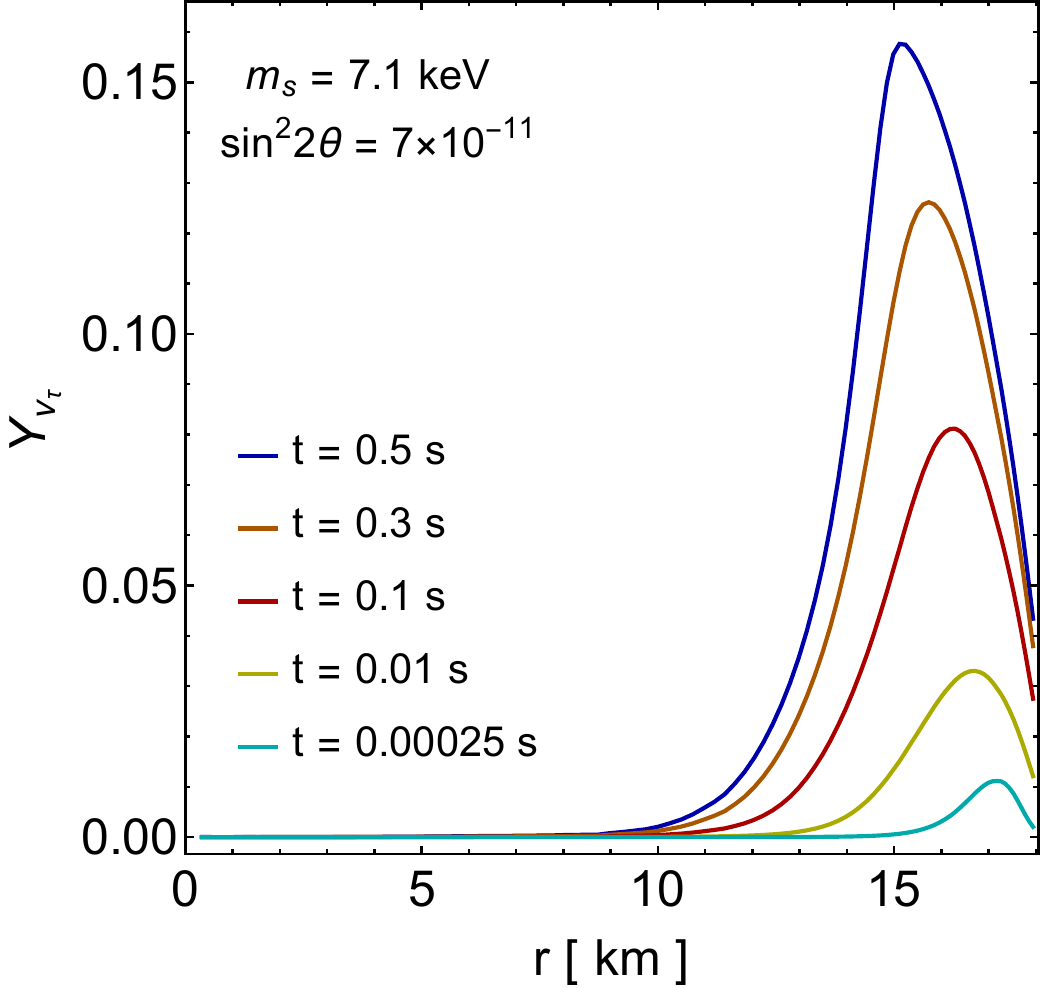}	
		\hspace*{0.5 cm}
		\includegraphics[width=0.3\textwidth]{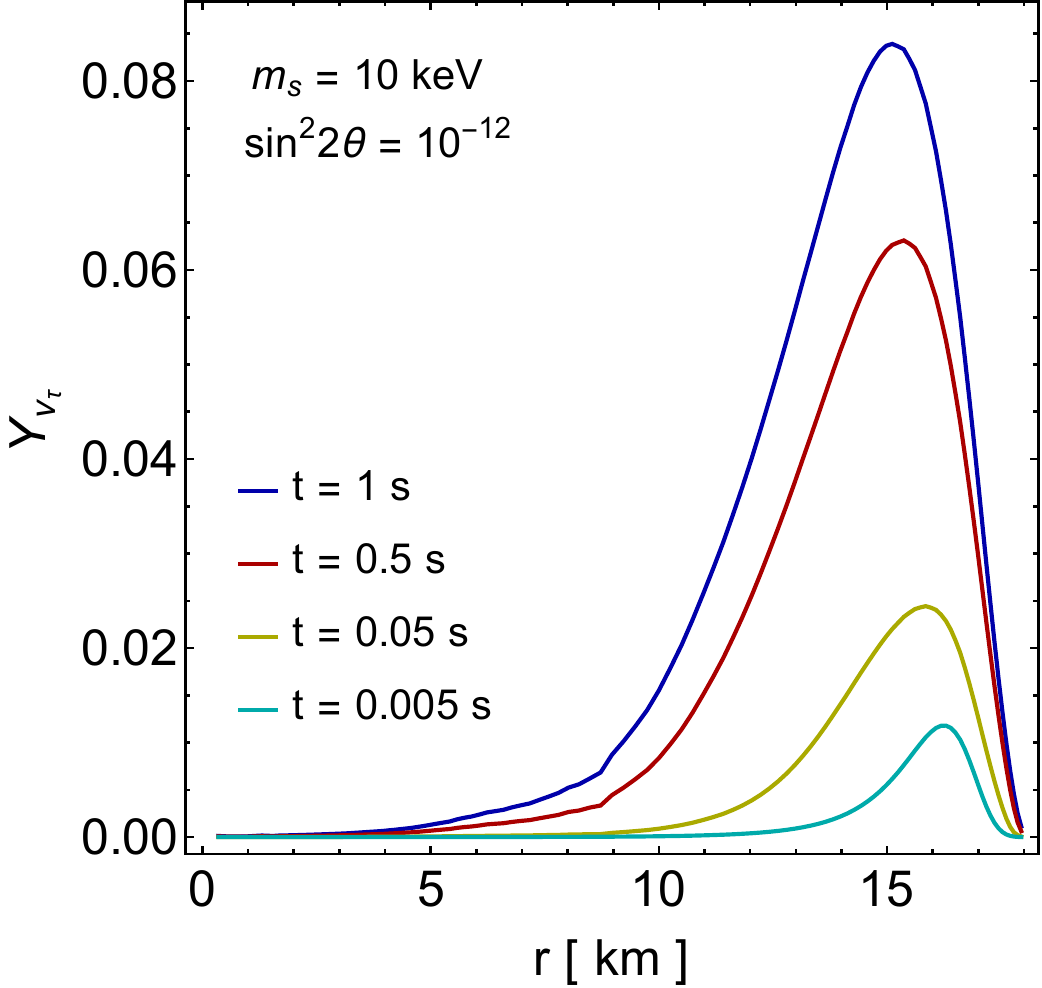}
		\hspace*{0.5 cm}
		\includegraphics[width=0.3\textwidth]{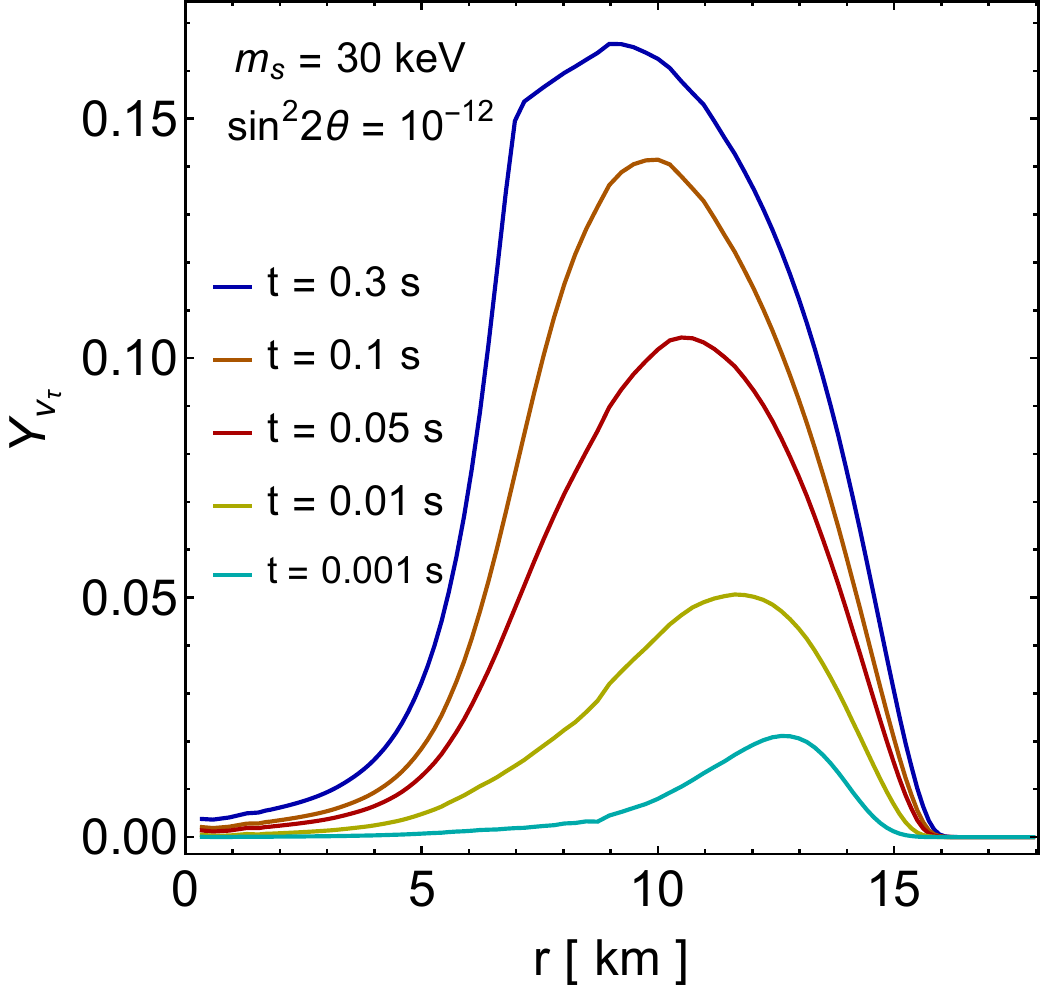}\bigskip\\ 
		\includegraphics[width=0.3\textwidth]{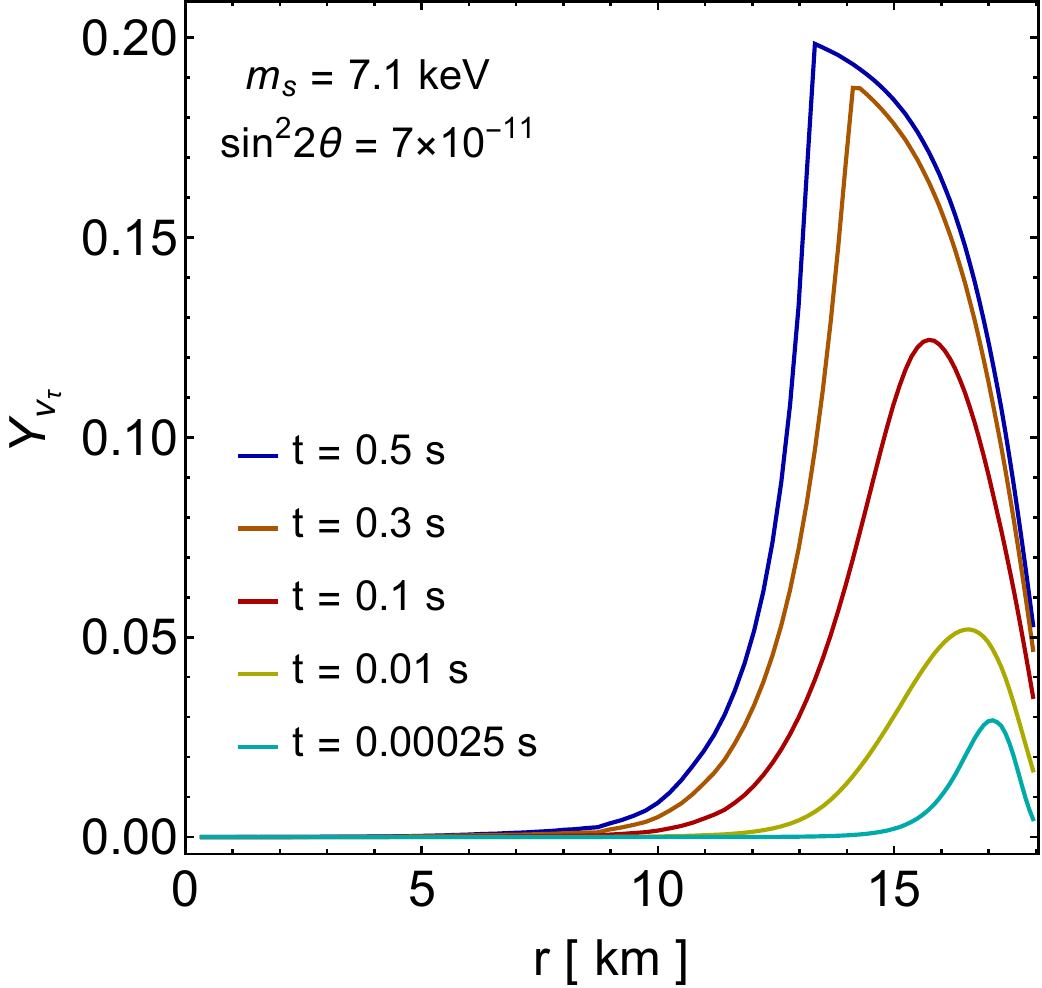}	
		\hspace*{0.5 cm}
		\includegraphics[width=0.3\textwidth]{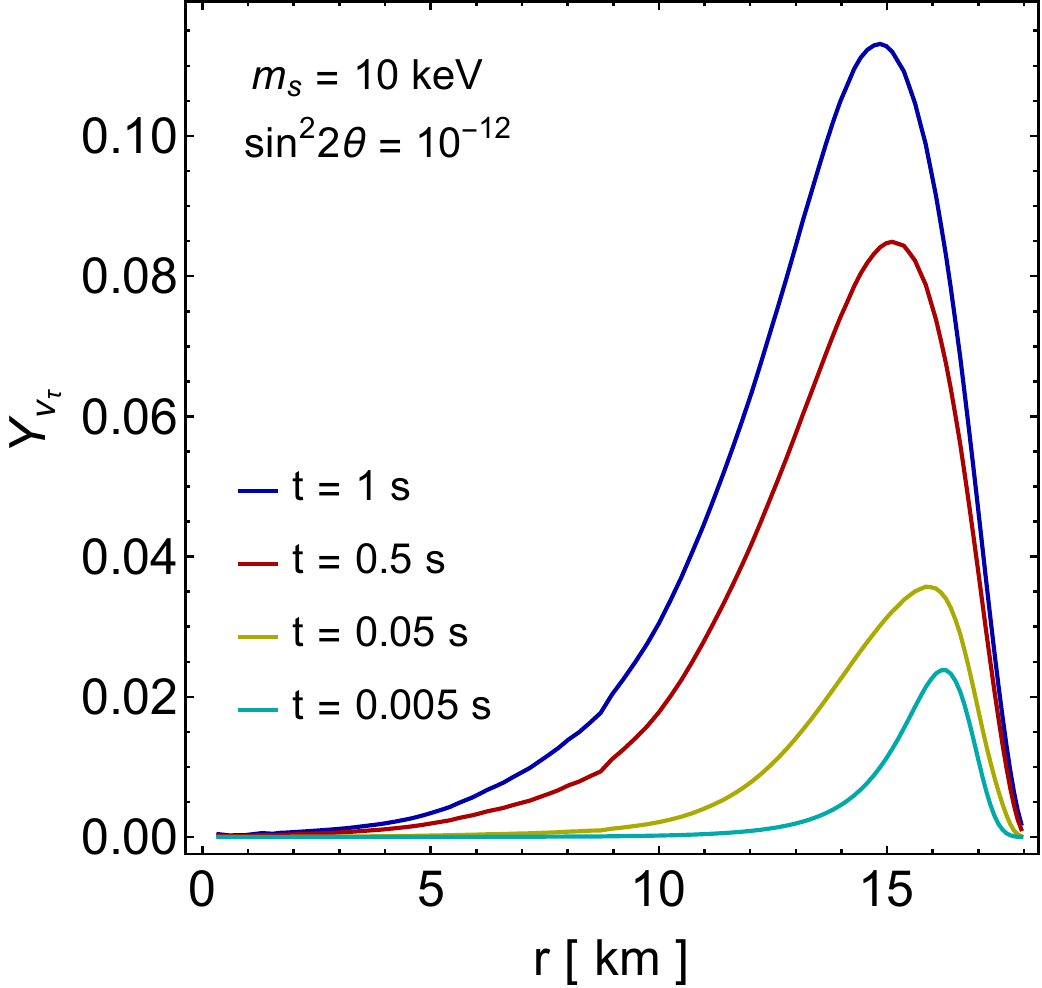}
		\hspace*{0.5 cm}
		\includegraphics[width=0.3\textwidth]{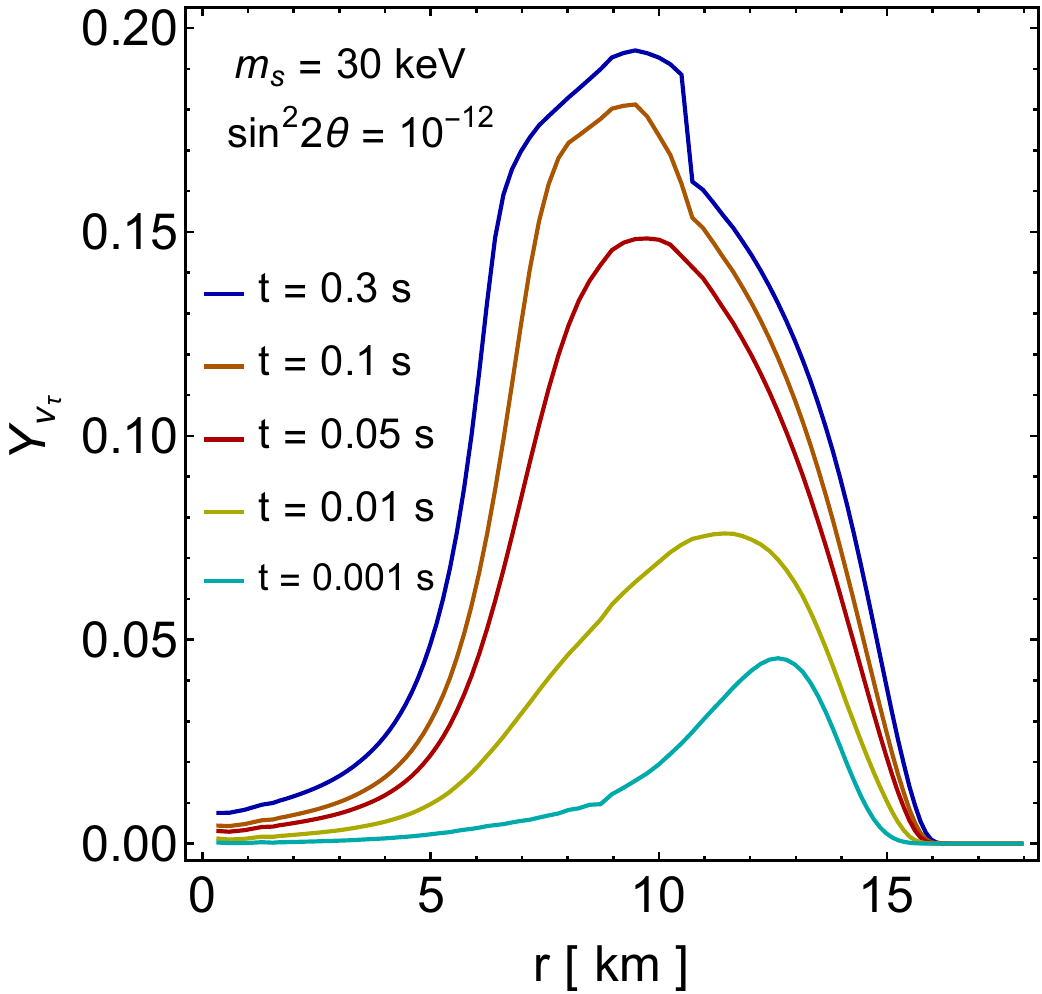}
		\caption{Radial profile of $Y_{\nu_\tau}$ at different times calculated for the indicated mixing parameters and for the protoneutron star conditions shown in Fig.~\ref{fig:pns}. The top and bottom panels differ in the adopted rate of change of $Y_{\nu_\tau}$ due to the MSW effect, with $\dot Y_{\nu_\tau}^{\rm MSW}$ and $\dot Y_{\nu_\tau}^{\rm MSW'}$ used for the former and latter, respectively.}
		\label{fig:profile}
	\end{figure*}
	
	\begin{figure*}[!t]
		\centering
		\includegraphics[width=0.3\textwidth]{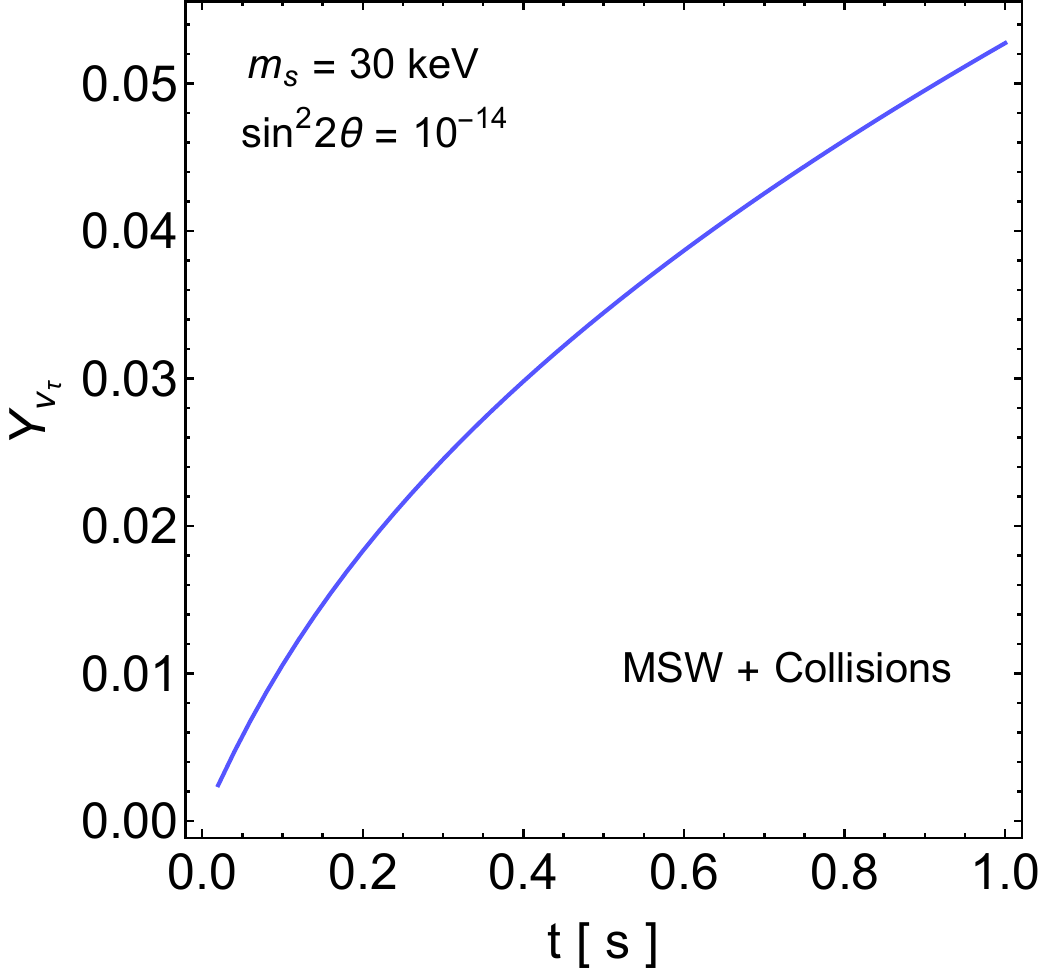}	
		\hspace*{0.5 cm}
		\includegraphics[width=0.3\textwidth]{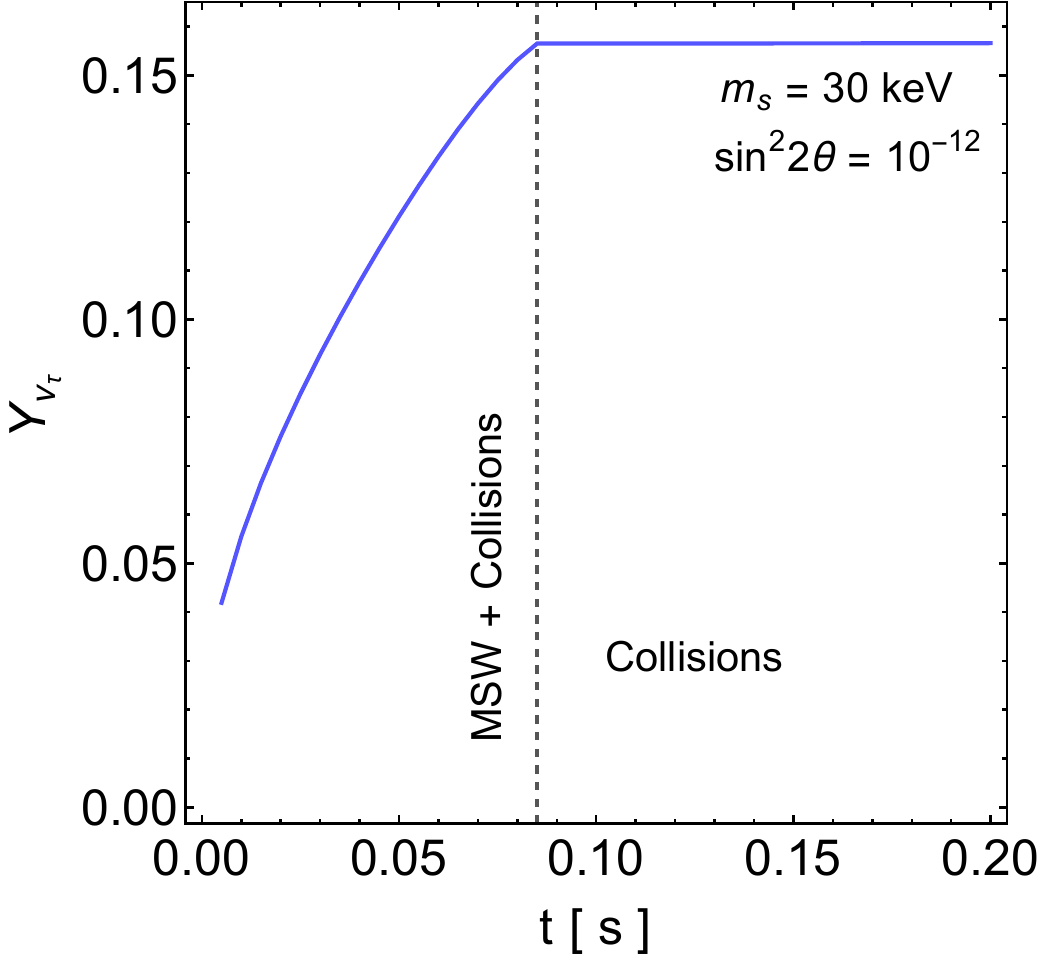}
		\hspace*{0.5 cm}
		\includegraphics[width=0.3\textwidth]{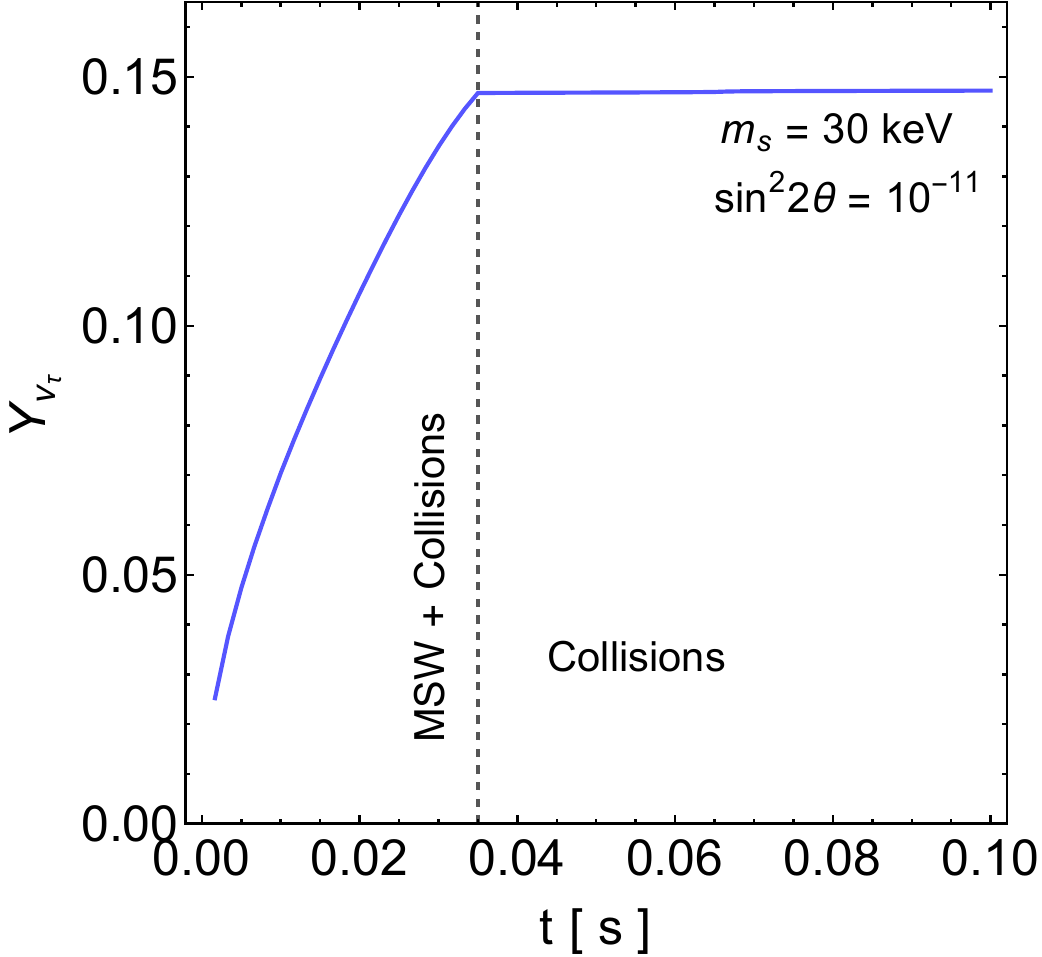}\bigskip\\ 
		\includegraphics[width=0.29\textwidth]{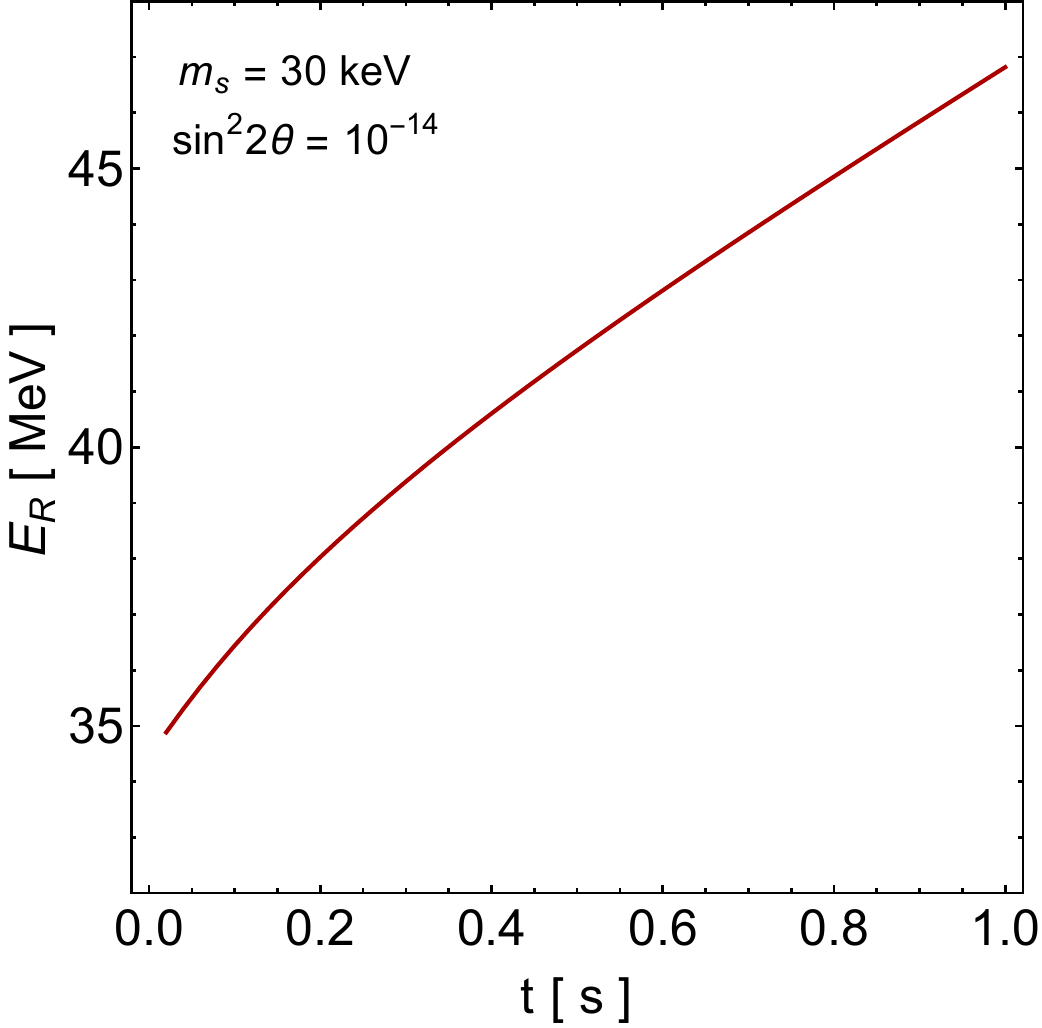}	
		\hspace*{0.5 cm}
		\includegraphics[width=0.3\textwidth]{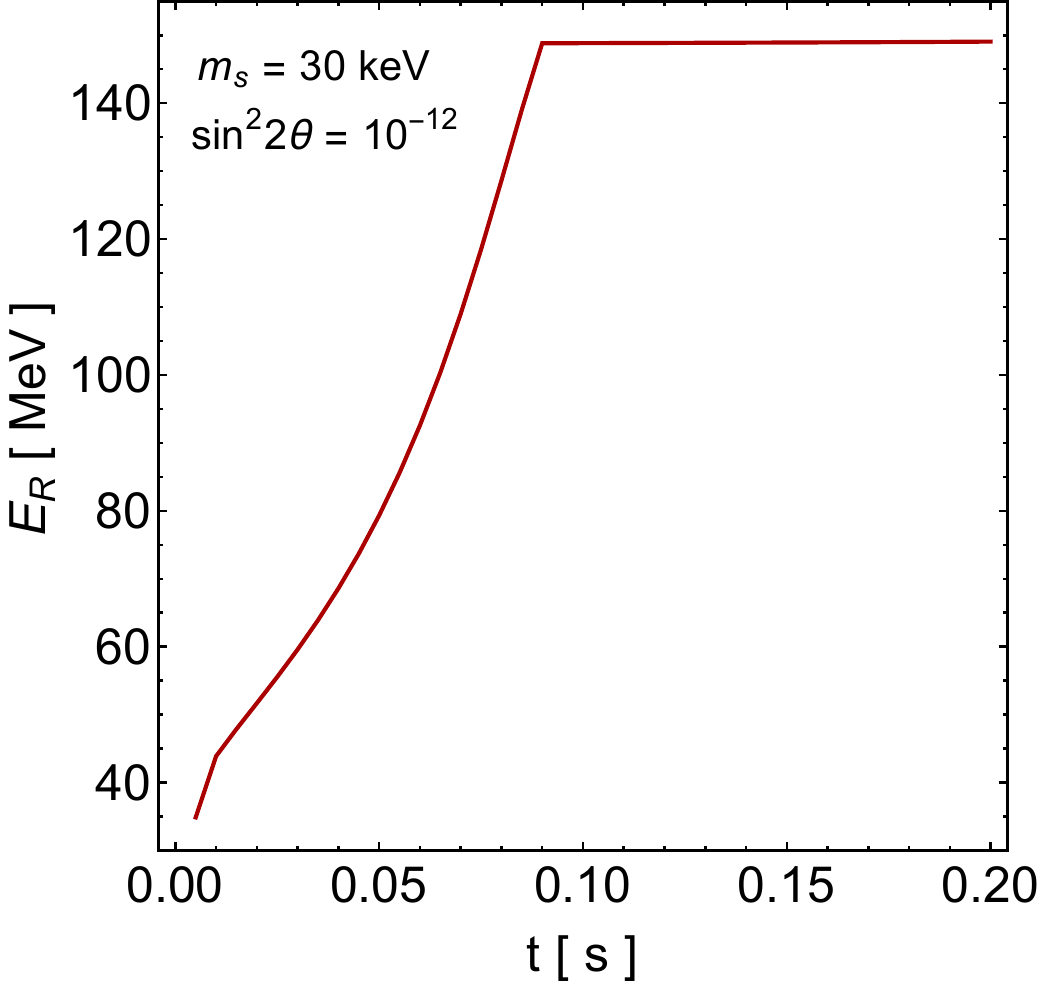}
		\hspace*{0.5 cm}
		\includegraphics[width=0.3\textwidth]{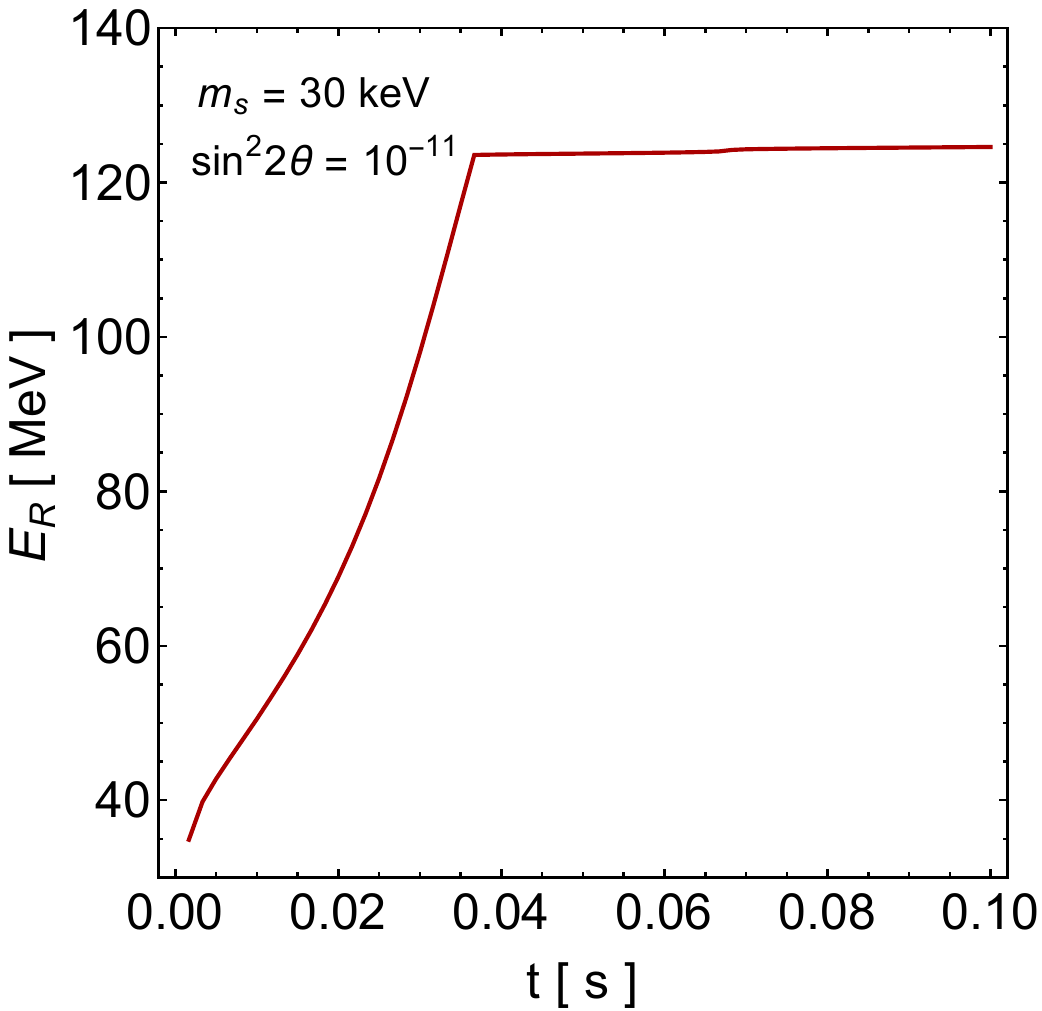}
		\caption{Time evolution of $Y_{\nu_\tau}$ (top panels) and the resonance energy $E_R$ (bottom panels) for a zone at $r \approx 8$~km for the indicated mixing parameters ($\dot Y_{\nu_\tau}^{\rm MSW}$ is used).}
		\label{fig:zone}
	\end{figure*}
	
	\section{Results}
	\label{sec:example}
	As discussed in Section~\ref{sec:setup}, treatment of $\nu_\tau$-$\nu_s$ and $\bar\nu_\tau$-$\bar\nu_s$ mixing includes production of
	$\nu_s$ and $\bar\nu_s$ through the MSW effect and collisions, as well as diffusion of the resulting $Y_{\nu_\tau}$. In turn,
	the evolution of $Y_{\nu_\tau}$ due to these processes changes the potentials $V_\nu$ and $V_{\bar\nu}$ for such mixing. Therefore, it is crucial to take this feedback into
	account in exploring the effects of such mixing. For illustration, we calculate the evolution of $Y_{\nu_\tau}$ for 
	$(m_s/{\rm keV},\sin^22\theta)=(7.1,7\times 10^{-11})$, $(10,10^{-12})$, and $(30,10^{-12})$, respectively, where 
	$m_s\approx\sqrt{\delta m^2}$ is the mass of the vacuum mass eigenstate almost coincident with the sterile neutrino. 
	The first set of parameters is suggested by interpreting the X-ray line emission near 3.55 keV from galaxy clusters as due to sterile neutrino decay~\cite{Bulbul:2014sua,Boyarsky:2014jta}, while the second set is chosen for sterile neutrinos to make up all the dark matter~\cite{Ng:2019gch,Roach:2022lgo}. For $m_s = 30$ keV, sterile neutrinos can make up all the dark matter for $\sin^2 2\theta \lesssim 10^{-14}$. Because the decay rate is proportional to $\sin^22\theta$, limits imposed by decay X-ray emission imply that sterile neutrinos can account for only $\lesssim 1\%$ of the dark matter for the third set of parameters~\cite{Ng:2019gch,Roach:2022lgo}.
	
	For the protoneutron star conditions, we take a snapshot of a $20\,M_\odot$ supernova model (with the SFHo nuclear equation of state) at 1 s post core bounce~\cite{Bollig2016,Mirizzi:2015eza,Bollig:2017lki,Bollig:2020xdr}. We assume that
	the radial profiles of $\rho$, $T$, $Y_e$, and $Y_{\nu_e}$ (see Fig.~\ref{fig:pns}) remain static during our calculation. The evolution of $Y_{\nu_\tau}(r,t)$ for
	a specific radial zone is determined by
	\begin{align}
		\frac{\partial Y_{\nu_\tau}}{\partial t}=\dot Y_{\nu_\tau}^{\rm MSW}+\dot Y_{\nu_\tau}^{\rm coll}+\dot Y_{\nu_\tau}^{\rm diff}.
		\label{eq:totalchange}
	\end{align}
	Using appropriate time steps, we obtain the radial profile of $Y_{\nu_\tau}$ at different times from the above equation. Because this equation
	is valid when active neutrinos, especially $\nu_\tau$ and $\bar\nu_\tau$, are in the diffusion regime, the radial 
	profile of interest is up to $r=R_\nu$. We have checked that our results are accurate. For example, the result for $t=0.3$~s shown in the top right panel of Fig.~\ref{fig:profile} has a maximum change of only $\approx 0.1\%$ when the resolution is doubled.
	
	Using the three sets of mixing parameters given above, we show in Fig.~\ref{fig:profile} the radial profile of $Y_{\nu_\tau}$ at different times ($t=0$ at the start of our calculation). The top and bottom panels differ in the adopted rate of change of $Y_{\nu_\tau}$ due to the MSW effect, with $\dot Y_{\nu_\tau}^{\rm MSW}$ [Eq.~(\ref{eq:rymsw})] and $\dot Y_{\nu_\tau}^{\rm MSW'}$ [Eq.~(\ref{eq:oldrymsw})] used for the former and latter, respectively. By the end of our calculation, the $Y_{\nu_\tau}$ profile is no longer changing significantly. In all cases, the calculation stops at or before $t=1$~s, when the assumption of static protoneutron star conditions starts to break down. We find that for mixing parameters consistent with current interpretation of and constraints on sterile neutrino dark matter, $\nu_\tau$-$\nu_s$ and $\bar\nu_\tau$-$\bar\nu_s$ mixing can produce $Y_{\nu_\tau}$ up to $\sim 0.1$--0.2 in protoneutron stars. Our results qualitatively resemble those of Refs.~\cite{2019JCAP...12..019S,PhysRevD.106.015017}. Despite the differences in the adopted protoneutron star conditions and in the approximations used, these results are also similar to each other quantitatively.
	
	In general, each $Y_{\nu_\tau}$ profile shown in Fig.~\ref{fig:profile} peaks at a certain radius. This feature occurs because production of $Y_{\nu_\tau}$ is dominated by escape of $\bar\nu_s$ converted from $\bar\nu_\tau$ through the MSW effect and this process is the most efficient at the radius where the local $\bar\nu_\tau$ energy distribution peaks near the energy for the local MSW resonance. As $Y_{\nu_\tau}$ increases with time, the potential $V_{\bar\nu}$ decreases and for a fixed $m_s$ the local resonance energy $E_R\approx m_s^2/(2V_{\bar\nu})$ increases with time. To match a larger $E_R$ with the peak of the local $\bar\nu_\tau$ energy distribution, the peak of the $Y_{\nu_\tau}$ profile shifts to smaller radii where the temperature is higher ($T$ decreases monotonically with radius for $r\gtrsim 8$~km as shown in Fig.~\ref{fig:pns}). Similarly, the resonance energy increases with $m_s$, and so the peak of the $Y_{\nu_\tau}$ profile tends to be at smaller radii for larger $m_s$.
	
	Note that although $\dot Y_{\nu_\tau}^{\rm MSW}$ and $\dot Y_{\nu_\tau}^{\rm MSW'}$ differ by a factor of $\sim 4$, the corresponding results on the $Y_{\nu_\tau}$ profile are rather close, which can be attributed to regulation by the feedback (see Section~\ref{sec:feedback}). Note also that the transition between $\bar\nu_s$ production through the MSW effect and that through collisions is assumed to occur at $\lambda_R=\delta r$. This abrupt transition leads to the sharp features in the $Y_{\nu_\tau}$ profile, which occur near the peak for $t=0.3$~s in the top and bottom right panels of Fig.~\ref{fig:profile} and at the peak for $t=0.3$ and 0.5~s in the bottom left panel. As mentioned in Sec.~\ref{sec:diff}, diffusion is inefficient, and therefore, it cannot smooth out these sharp features.

	\section{Feedback and evolution of tau-neutrino lepton number}
	\label{sec:feedback}
	To elaborate on the feedback of $\nu_{\tau}$-$\nu_s$ and $\bar\nu_{\tau}$-$\bar\nu_s$ mixing in protoneutron stars, we focus on the evolution of $Y_{\nu_\tau}$ in a specific radial zone. We choose a zone at $r\approx 8$~km with $\rho\approx 4\times 10^{14}$~g~cm$^{-3}$, $T\approx58$~MeV, $Y_e\approx0.18$, and $Y_{\nu_e}\approx0.013$. We take $m_s = 30$~keV along with $\sin^22\theta = 10^{-14}$, $10^{-12}$, and $10^{-11}$, respectively, and show in Fig.~\ref{fig:zone} the $Y_{\nu_\tau}$ in this zone as a function of time ($\dot Y_{\nu_\tau}^{\rm MSW}$ is used). In all cases, $Y_{\nu_\tau}$ initially increases mainly due to escape of $\bar\nu_s$ converted from $\bar\nu_\tau$ through the MSW effect. This increase of $Y_{\nu_\tau}$ decreases the potential $V_{\bar\nu}$ and hence increases the local resonance energy $E_R$ (see Fig.~\ref{fig:zone}). Consequently, the mean free path $\lambda_R$ at this energy decreases. Meanwhile, the radial width $\delta r$ of the resonance region tends to increase due to the flattening of the $V_{\bar\nu}$ profile across the zone. When $\lambda_R$ falls below $\delta r$, $Y_{\nu_\tau}$ production due to the MSW effect turns off and $Y_{\nu_\tau}$ essentially stops evolving because the rate of production through collisions is small and diffusion is inefficient. This turnoff is clearly seen in Fig.~\ref{fig:zone} for $\sin^22\theta = 10^{-12}$ and $10^{-11}$. It occurs earlier for the larger $\sin^22\theta$ because $\delta r$ is proportional to $\sin 2 \theta$. For $\sin^22\theta = 10^{-14}$, $\delta r$ is sufficiently small and never exceeds $\lambda_R$. Consequently, $Y_{\nu_\tau}$ production due to the MSW effect persists throughout our calculation in this case. However, this small $\sin^22\theta$ also renders the MSW conversion of $\bar\nu_\tau$ into $\bar\nu_s$ highly inefficient [see Eqs.~(\ref{eq:LZ}) and (\ref{eq:gamma})]. Consequently, although our calculation is carried out until $t=1$~s for this case, the final $Y_{\nu_\tau}$ is still a factor of $\approx 3$ smaller than those obtained at $t\approx0.085$ and 0.035~s for $\sin^22\theta = 10^{-12}$ and $10^{-11}$, respectively.
	
	In addition to the turnoff for $\lambda_R<\delta r$ discussed above and indicated by the Heaviside step function in Eq.~(\ref{eq:rymsw2}), $\dot Y_{\nu_\tau}^{\rm MSW}$ is regulated approximately by the factor $(E_R^3e^{-E_R/T})e^{-\eta_{\nu_\tau}}$, which peaks at $E_R=3T$. For the selected zone with $T\approx 58$~MeV, $E_R$ is always below $3T$ (see Fig.~\ref{fig:zone}) and this factor does not change the above discussion of this zone qualitatively. However, it could effectively turn off the MSW production of $Y_{\nu_\tau}$ for other zones even when $\lambda_R$ still exceeds $\delta r$. This effective turnoff can occur when $E_R$ increases beyond $3T$ and hence $\dot Y_{\nu_\tau}^{\rm MSW}$ is suppressed exponentially. This suppression is further enhanced by the increase of $\eta_{\nu_\tau}$ with $Y_{\nu_\tau}$. Therefore, the feedback of $\nu_{\tau}$-$\nu_s$ and $\bar\nu_{\tau}$-$\bar\nu_s$ mixing in protoneutron stars always tends to turn off the effects of such mixing.
	
	\section{Discussion and conclusions}
	\label{sec:discuss}
	We have presented a concise and analytical treatment of the processes associated with $\nu_\tau$-$\nu_s$ and $\bar\nu_\tau$-$\bar\nu_s$ mixing in protoneutron stars. These processes include sterile neutrino production through the MSW effect and collisions as well as evolution of the $\nu_\tau$ lepton number fraction $Y_{\nu_\tau}$ through escape of sterile neutrinos and diffusion. We find that for mixing parameters consistent with current interpretation of and constraints on sterile neutrino dark matter of ${\cal{O}}(10)$~keV in mass, $\nu_\tau$-$\nu_s$ and $\bar\nu_\tau$-$\bar\nu_s$ mixing can produce $Y_{\nu_\tau}$ up to $\sim0.1$--0.2 in protoneutron stars. This result is consistent with previous studies \cite{2019JCAP...12..019S,PhysRevD.106.015017}. In addition, we find that evolution of $Y_{\nu_\tau}$ is dominated by MSW conversion of $\bar\nu_\tau$ into $\bar\nu_s$, and that both sterile neutrino production through collisions and $Y_{\nu_\tau}$ diffusion are inefficient. Mainly through feedback on the potential for $\bar\nu_\tau$-$\bar\nu_s$ conversion, the increase of $Y_{\nu_\tau}$ eventually turns off $\bar\nu_s$ production through the MSW effect. Therefore, we confirm conclusions of previous studies \cite{2011PhRvD..83i3014R,2019JCAP...12..019S} that feedback of $\nu_\tau$-$\nu_s$ and $\bar\nu_\tau$-$\bar\nu_s$ mixing tends to turn off the effects of such mixing in protoneutron stars.
	
	In this work, we have made some approximations and assumptions that appear reasonable but are worth discussing. The main approximation is that we have taken radially-propagating neutrinos as representative of flavor evolution governed by the MSW effect in deriving $\dot Y_{\nu_\tau}^{\rm MSW}$. As discussed in Section~\ref{sec:MSW}, a full treatment should account for all the directional dependence of $\bar\nu_s$ production through the MSW effect. Another approximation is the sharp transition between $\bar\nu_s$ production through the MSW effect and that through collisions at $\lambda_R=\delta r$. Both the above approximations can be improved by results from future numerical simulations of $\bar\nu_\tau$-$\bar\nu_s$ mixing for an isotropic $\bar\nu_\tau$ source in a box of linear size comparable to typical $\bar\nu_\tau$ mean free path, where collisions occur in a medium of varying density.
	
	The major assumption of this work is that $\nu_\tau$-$\nu_s$ and $\bar\nu_\tau$-$\bar\nu_s$ mixing only affects $\nu_\tau$ and $\bar\nu_\tau$ in protoneutron stars while all the other conditions are unchanged and remain static for $\sim 1$~s. A self-consistent and full treatment should calculate the dynamical evolution of a protoneutron star by including energy loss through sterile neutrinos and taking into account the modification of $\nu_\tau$ and $\bar\nu_\tau$ distributions by the $\nu_\tau$ lepton number created through such mixing. Only such a calculation can definitively constrain the mixing parameters associated with sterile neutrino dark matter of ${\cal{O}}(10)$~keV in mass.
	
\section*{Acknowledgments} 
	We thank George Fuller, Georg Raffelt, Anna Suliga, Irene Tamborra, and Meng-Ru Wu for helpful discussion. We are also indebted to Daniel Kresse and Thomas Janka for providing the supernova model used in this work. This work is supported in part by the National Science Foundation (Grant No. PHY-2020275), the Heising-Simons Foundation (Grant 2017-228), and the U.S. Department of Energy (Grant No. DE-FG02-87ER40328).
	
\bibliographystyle{JHEP}
\bibliography{references}
\end{document}